\documentclass[12pt]{article}
\pdfoutput=1
\usepackage{graphicx}
\usepackage{latexsym}
\usepackage{ifpdf}
\ifpdf
\usepackage{textcomp}
\usepackage{amssymb}

\usepackage{amsfonts,amsthm,amstext,amscd}
\usepackage{slashed}
\usepackage{amsfonts,amssymb,amsthm,amstext,amscd}

\usepackage{amsmath}
\newcommand{\beqa}{\begin{eqnarray}}
\newcommand{\eeqa}{\end{eqnarray}}

\def\d{{\rm d}}

\newcommand{\be}{\begin{equation}}
\newcommand{\ee}{\end{equation}}
\newcommand{\beq}{\begin{equation}}
\newcommand{\eeq}{\end{equation}}
\newcommand{\bea}{\begin{eqnarray}}
\newcommand{\eea}{\end{eqnarray}}

\newcommand{\bear}{\begin{eqnarray}}
\newcommand{\eear}{\end{eqnarray}}
\begin{document}
\baselineskip=15.5pt
\pagestyle{plain}
\setcounter{page}{1}
%--------+---------+---------+---------+---------+---------+---------+
%Body

\def\r{\rho}
\def\CC{{\mathchoice
{\rm C\mkern-8mu\vrule height1.45ex depth-.05ex
width.05em\mkern9mu\kern-.05em}
{\rm C\mkern-8mu\vrule height1.45ex depth-.05ex
width.05em\mkern9mu\kern-.05em}
{\rm C\mkern-8mu\vrule height1ex depth-.07ex
width.035em\mkern9mu\kern-.035em}
{\rm C\mkern-8mu\vrule height.65ex depth-.1ex
width.025em\mkern8mu\kern-.025em}}}

% FONTS
\newfont{\namefont}{cmr10}
\newfont{\addfont}{cmti7 scaled 1440}
\newfont{\boldmathfont}{cmbx10}
\newfont{\headfontb}{cmbx10 scaled 1728}
\renewcommand{\theequation}{{\rm\thesection.\arabic{equation}}}

\par\hfill{IFUP-TH-2014-10}
%\par\hfill{}
\vspace{1cm}

\begin{center}
{\huge{\bf Holographic QCD with Dynamical Flavors
}}
\end{center}

\vskip 10pt

\begin{center}
{\large Francesco Bigazzi$^{a}$ and  Aldo L. Cotrone$^{b}$}
\end{center}

\vskip 10pt
\begin{center}
\textit{$^a$ INFN, Sezione di Pisa; Largo B. Pontecorvo 3, I-56127 Pisa, Italy.}\\
\textit{$^b$ Dipartimento di Fisica e Astronomia, Universit\`a di
Firenze and INFN, Sezione di Firenze; Via G. Sansone 1, I-50019 Sesto Fiorentino
(Firenze), Italy.}\\

\vskip 10pt
{\small fbigazzi@pi.infn.it, cotrone@fi.infn.it}
\end{center}

\vspace{25pt}

\begin{center}
 \textbf{Abstract}
\end{center}
\noindent
Gravity solutions describing the Witten-Sakai-Sugimoto model of holographic QCD with dynamical flavors are presented. 
The field theory is studied in the Veneziano limit, at first order in the ratio of the number of flavors and colors. 
The gravity solutions are analytic and dual to the field theory either in the confined, low temperature phase or in the deconfined, high temperature phase with small baryonic charge density. 
The phase diagram and the flavor contributions to vacuum (e.g. string tension and hadron masses) and thermodynamical properties of the dual field theory are then deduced. 
The phase diagram of the model at finite temperature and imaginary chemical potential, as well as that of the unflavored theory at finite $\theta$ angle are also discussed in turn, showing qualitative similarities with recent lattice studies. Interesting degrees of freedom in each phase are discussed.\\
Covariant counterterms for the Witten-Sakai-Sugimoto model are provided both in the probe approximation and in the backreacted case, allowing for a standard holographic renormalization of the theory. 
 
\newpage

%%%%%%%%%%%%%%%%%%%%%%%%%%%%%%%%%%%%%%%%%%%%%%%%%%%%%%
\section{Introduction}

The Witten-Sakai-Sugimoto (WSS) model \cite{witten,ss} is the most studied top-down example of ``holographic QCD''.
Its realization in string theory describes, in the low energy limit, a four dimensional $SU(N_c)$ non-supersymmetric Yang-Mills theory coupled to adjoint massive matter and $N_f$ chiral massless fermions - quarks in the fundamental representation.
Its popularity is due to its ability in describing in a very simple and calculable way many physical properties interesting for QCD, such as confinement, chiral symmetry breaking, confinement-deconfinement transition, etc.\footnote{The literature on the model is huge, so it is impossible to provide here an exhaustive list of references.}
In fact, the WSS model is the holographic theory closest to (planar) QCD,\footnote{Of course this is only true unless one abandons the controllable framework of top-down constructions.} although it is not its precise dual. 

In the original form of the Witten-Sakai-Sugimoto model the quarks in the fundamental representation are quenched, non-dynamical degrees of freedom. 
In this paper we construct holographic solutions corresponding to dynamical fundamental matter, at first order in the Veneziano limit.
That is, the gravitational solutions we present provide an accurate description of the field theory in the limit of large number of colors and flavors, $N_c\gg1, N_f \gg 1$, with fixed ratio $\epsilon_f \sim N_f/N_c$, at first order in the small $\epsilon_f$-expansion.
We study the model both in the confined and deconfined phase (below and above a critical temperature) and we also include a finite, small baryonic charge density.
Our solutions describe a configuration where the flavor symmetry is completely Abelian - $U(1)^{N_f}_L \times U(1)^{N_f}_R$ prior to chiral symmetry breaking.

The gravity solutions allow for the study of effects of dynamical flavors on the observables of the theory.
As illustrative examples, we first briefly consider the string tension and the mass spectrum of some hadrons in the confined case.
Then, we re-derive the thermodynamics of the solutions (previously studied in the quenched limit in \cite{Aharony:2006da, Kim:2006gp, bergman}) and discuss the phase diagrams (in a section hopefully accessible to the non expert on holography).
Our aim is to compare the holographic results to those obtained in lattice QCD. 
We first consider the flavor dependence of the deconfinement temperature\footnote{Just as large $N_c$ QCD, the WSS model in the small $N_f/N_c$ Veneziano limit experiences a first order transition between the confined and the deconfined phase. Thus there is a well defined critical temperature. In real world QCD, instead, the transition is replaced by a sharp crossover.} and its behavior as a function of the baryon chemical potential, obtaining qualitative matching with lattice QCD results. 
Then we focus on the phase diagram of the model at finite temperature and imaginary chemical potential, as well as on that of the unflavored theory at finite $\theta$ angle. 
The holographic model is observed to provide a natural realization of a qualitative ``duality'' between these two phase diagrams, recently observed in lattice (quenched) QCD \cite{massimo1}.
In the latter paper, in fact, the phase diagram at finite temperature and $\theta$ angle is observed to have the same qualitative features as an ``inverted'' (Roberge-Weiss) phase diagram at finite temperature and imaginary chemical potential.
In the holographic model this is mainly due to a symmetry of the background under the exchange of two circles in the geometry.
The holographic model also allows to strengthen a natural proposal for the relevant degrees of freedom in the two phases: instantons in the deconfined phase and baryons in the confined one.

Coming to some technical details, we construct the solutions by making use of the smearing technique \cite{Bigazzi:2005md,Casero:2006pt} (see also \cite{Benini:2006hh} and \cite{Nunez:2010sf,Bigazzi:2011db} for reviews).
The WSS model consists in adding ``flavor'' $D$8-branes to the non-supersymmetric background generated by $D4$-branes wrapped on a circle \cite{witten}. This background is dual, in the low energy limit, to a Yang-Mills theory with adjoint massive matter.
The $D8$-branes, which describe the flavor sector, are placed at a fixed value of the wrapped circle mentioned above \cite{ss}.
Studying their backreaction on the background in the localized case is a daunting task and it has been performed only in a particular limit in \cite{sonne}. Notice that an added difficulty in the WSS model, with respect to other setups based e.g. on $D3-D7$ branes, is that supersymmetry is completely broken.

We are able to take into account the backreaction of the $D8$-branes, and so to consider the dynamics of fundamental matter, by homogeneously smearing a large number of $D8$-branes on the wrapped circle,\footnote{The smearing selects the Abelian form of the flavor symmetry.} 
such that the isometries of the original background are preserved. 
This configuration greatly simplifies the setting, reducing the equations to be solved to ODEs instead of PDEs.
The system is still extremely complicated: it is non-supersymmetric - giving a set of coupled second order equations - and the finite charge density implies the presence of highly non-linear terms.
Nevertheless, surprisingly enough, the system admits analytic solutions in the limit of small flavor backreaction and small charge density.
The parameters of the solutions can be chosen in such a way that the backgrounds are completely regular in the dual IR regime.
In the dual UV regime, on the contrary, they present a non-removable divergence, due to the presence of a Landau pole in field theory.\footnote{The unflavored solution is dual to a conformal theory in the UV \cite{witten}.}
The latter is reflected in the holographic running coupling too.

Last but not least, in order to compute the free energy of the system, we holographically renormalize the theory.
As far as we know, despite a huge number of studies of the WSS model, its holographic renormalization is not known in a covariant way so far.
Thus, we provide covariant counterterms for the WSS model, filling the gap in the literature.
The form of the counterterms is the same both in the probe approximation and in the backreacted case (at leading order in $\epsilon_f$).

The paper is organized as follows.
In the next section we provide a very short review of the WSS model in the quenched case.
Then, in section \ref{secconfinedcase} we construct the gravitational solution with backreacting $D8$-branes in the dual confined phase.
Some physical quantities, such as the running coupling, string tension and the baryon vertex and vector meson mass spectra, are briefly discussed as well.
The holographic renormalization of the WSS model is described in section \ref{holren}.  
In section  \ref{secdeconfinedcase} we present the gravitational solution dual to the deconfined phase with finite charge density, and we derive its thermodynamics in section \ref{secthermo}.
Section \ref{secphases}, which is written also for non experts in holography, presents the discussion of the critical deconfinement temperature , the phase diagrams and some interesting degrees of freedom of the theory.
We conclude in section \ref{secconclusions} with some comments on the solution, its limitations and potential applications.
In the appendix we write the solution in the particular limit considered in \cite{Antonyan:2006vw}.

%%%%%%%%%%%%%%%%%%%%%%%%%%%%%%%%%%%%%%%%%%%%%%%%%%%%%%%%
\section{A review of  the Witten-Sakai-Sugimoto model}\label{secreview}
\setcounter{equation}{0}
In \cite{witten}, a large $N_c$ non-supersymmetric Yang-Mills theory in $3+1$ flat space-time dimensions was obtained by Witten as the low energy limit of a Kaluza-Klein (KK) reduction of a particular $4+1$ dimensional $SU(N_c)$ Yang-Mills theory coupled to massless adjoint scalar and fermionic matter fields. 
This theory describes the low energy dynamics of open strings whose end-points are attached (with Dirichlet boundary conditions) to a stack of $N_c$ parallel $D4$-branes ($4+1$-dimensional hyperplanes) placed in an ambient ten dimensional Minkowski space-time. 
The reduction is realized by compactifying the theory on a circle $S_{x_4}$ of length $\beta_4$ along the fourth space direction $x_4$ and choosing periodic (resp. anti-periodic) boundary conditions for bosons (resp. fermions). 
In this way the massless modes at energy $E\ll 1/\beta_4$ are those of a $3+1$ dimensional $SU(N_c)$ Yang-Mills theory. 
The other modes get masses $M_{KK}$ of the order of $1/\beta_4$. 
If we denote by $T_s$ the string tension of the confining $3+1$ dimensional Yang-Mills theory, the low energy theory can be decoupled from the Kaluza-Klein modes provided $T_s/M^2_{KK}\equiv 2\lambda_4/27 \pi \ll1$. Here, $\lambda_4$ is the 4d UV 't Hooft coupling.

Unfortunately, Witten's model in the most interesting $\lambda_4 \sim 1$ regime has no known simple description. 
However, we can obtain many detailed informations in the $\lambda_4\gg1$ regime, where the theory (in the large $N_c$ limit) is conjectured to have a holographic dual description in terms of a classical theory of gravity on a background which arises as the near-horizon limit of the one sourced by the $N_c$ $D4$-branes. 
This background has the topology of a product $\mathbb{R}^{1,3}\times \mathbb{R}_u \times S_{x_4}\times S^4$. 
Here $\mathbb{R}^{1,3}$ is the $3+1$ dimensional flat Minkowski space-time, while $\mathbb{R}_u$ denotes the real semi-axis spanned by a radial coordinate $u$ which is roughly the geometric counterpart of the Renormalization Group energy scale in the dual field theory. 
The $(u, x_4)$ subspace is a {\it cigar}, with the $S_{x_4}$ circle smoothly shrinking to zero size at a finite value $u_0$ of the radial coordinate.\footnote{It is this shrinking cycle which automatically implements the (anti)periodic boundary conditions on the dual fields along $S_{x_4}$.}  
Finally, $S^4$ is a compact four-dimensional sphere, whose isometry group $SO(5)$ is holographically mapped into a global symmetry group under which the massive Kaluza-Klein fields (which are not decoupled in the large $\lambda_4$ regime) rotate.
Despite being in a regime where the interesting Yang-Mills part is intrinsically coupled with spurious Kaluza-Klein modes, the $\lambda_4\gg1$ theory displays confinement and (once coupled to chiral massless quarks) chiral symmetry breaking at $T=0$. 

\begin{figure}[t]
\centering
\includegraphics[scale=0.5]{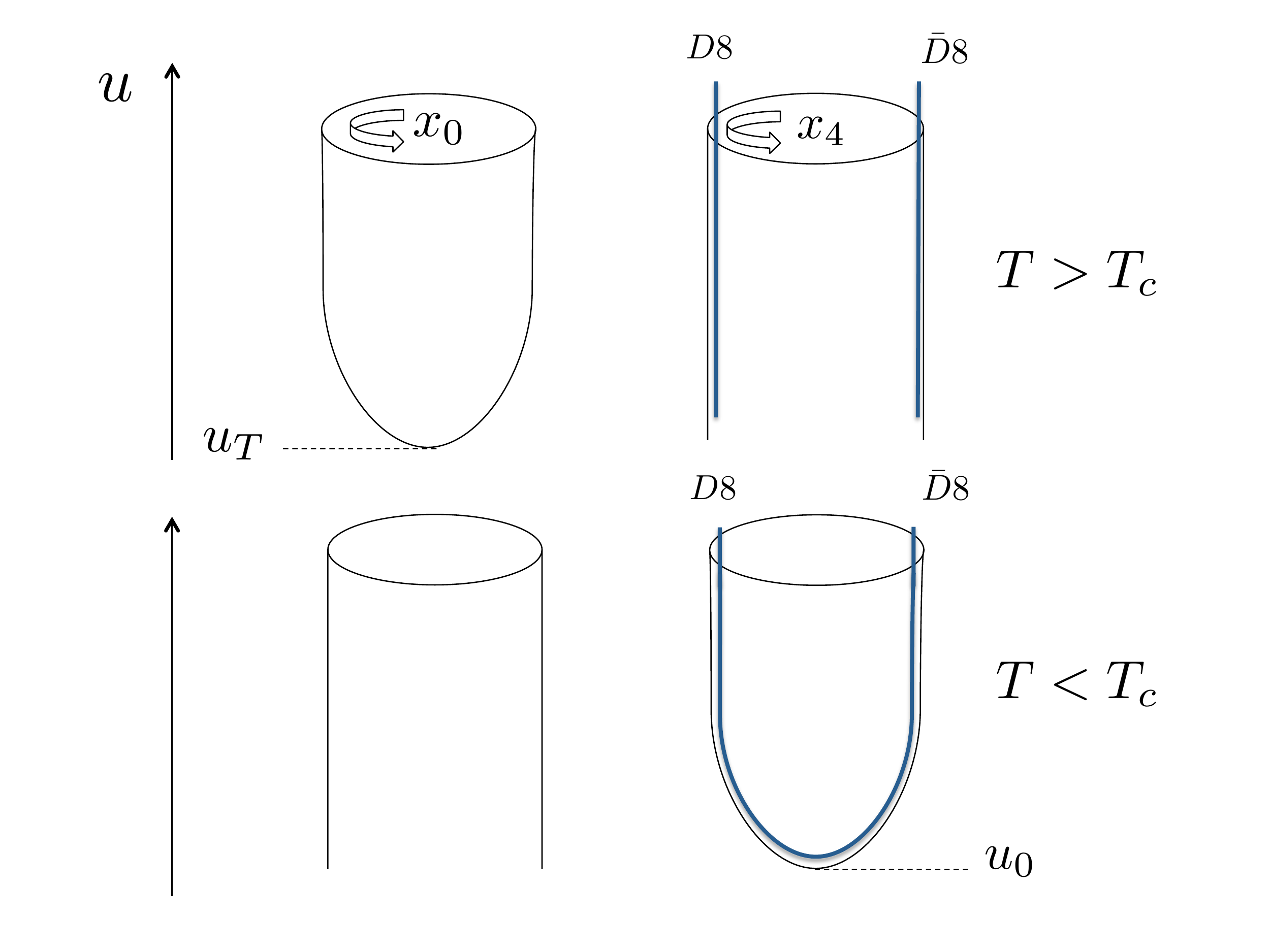}
\caption{\small{The cylinder and cigar sections of Witten-Sakai-Sugimoto's gravity solution at $T>T_c$ (deconfined phase) and $T<T_c$ (confined phase). 
Here $u$ is the radial coordinate (the geometric counterpart of  the RG energy scale in the dual field theory) and $x_0$ and $x_4$ are Euclidean time and space directions compactified on circles of length $1/T$ and $2\pi/M_{KK}$ respectively. 
The $N_f$ $D8$-branes (resp. ${\bar D}8$-branes) account for the presence of $N_f$ right-handed (resp. left-handed) massless quarks in the dual quantum field theory. 
These branes connect in a ``U''-shaped configuration at $T<T_c$: this provides a simple geometrical realization of chiral symmetry breaking. }}
\label{fig0}
\end{figure}

The introduction of  $N_f$ chiral fundamental massless quarks in Witten's model can be achieved by adding suitably embedded $N_f$ $D8-{\bar D}8$-branes (which are localized in the $x_4$ direction) in the background, as described by Sakai and Sugimoto in  \cite{ss}. 
In fact, quark fields arise in the massless spectrum of the open strings stretching between the ``color" $D4$-branes and the ``flavor" $D8$-branes. 
Neglecting the flavor brane backreaction on the background amounts on treating the quarks in the quenched approximation. 
The flavor branes support a $U(N_f)\times U(N_f)$ gauge group which is holographically mapped into the global flavor symmetry of the dual field theory. 
Chiral symmetry breaking is simply realized in the model: at $T=0$ the flavor branes will find energetically convenient to connect into a ``U''-shaped configuration, see figure \ref{fig0}. 
This will automatically break the original flavor symmetry group into a diagonal $SU(N_f)\times U(1)_B$ subgroup, where $U(1)_B$ is the baryon symmetry. The axial symmetry $U(1)_A$, instead, is broken by the anomaly.

Going to finite temperature amounts on compactifying the (Euclideanized) time direction $x_0$ on a circle $S_{x_0}$ of length $\beta=1/T$. There are two allowed gravity backgrounds where this is realized. 
A first one has the same structure as that at $T=0$. The $(x_0, u)$ subspace is a cylinder while the $(x_4, u)$ one is still a cigar. The full background is given by 
\bea\label{witten0}
&& ds^2 = \left(\frac{u}{R}\right)^{3/2}\left[dx_{\mu} dx^{\mu} + f(u) dx_4^2\right] + \left(\frac{u}{R}\right)^{-3/2}\left[\frac{du^2}{f(u)} + u^2 d\Omega_4^2\right]\,,\nonumber \\
&& e^{\phi}= g_s\left(\frac{u}{R}\right)^{3/4}\,,\qquad F_4 = \frac{3N_c}{4\pi}\omega_4\,,\qquad f(u) = 1- \frac{u^3_{0}}{u^3}\,.
\eea
Here $\mu=0,1,2,3$. $R$ measures the curvature radius of the background in string units, $R= (\pi g_s N_c)^{1/3} l_s$, where $g_s$ is the string coupling and $l_s$ is the string length. 
Moreover $d\Omega_4^2$ is the metric of the transverse $S^4$, $\phi$ is the dilaton, $F_4$ is a Ramond-Ramond four-form and $\omega_4$ is the volume form of $S^4$. 
The 't Hooft coupling of the 4d theory in the UV is given by $\lambda_4=g_{YM}^2N_c = 4\pi^2 g_s N_c l_s/\beta_4$. The shrinking of the $S_{x_4}$ circle at $u=u_{0}$ is evident from the fact that $f(u_{0})=0$. The $(u, x_4)$ subspace is a cigar, i.e. there are no conical singularities at $u=u_{0}$ provided  the relation
\be
9\beta_4^2 u_{0}= 16 \pi^2 R^3
\label{uomp}
\ee
holds. The $x_4$ periodicity $\beta_4$ is related to the mass scale $M_{KK}$ of the Kaluza-Klein modes by $\beta_4=2\pi/M_{KK}$.
The confining string tension is $T_{s}=u_0^{3/2}/2\pi \alpha' R^{3/2}=2\lambda_4 M_{KK}^2/27\pi$.

A second allowed solution (which thus has the same $u\rightarrow\infty$ asymptotic as the one above) is instead a black-hole background\footnote{It has been claimed in \cite{Mandal:2011ws} that the black hole solution is not smoothly connected to the deconfined phase of pure YM in 4d for $\lambda_4 \ll 1$. The ``correct'' background in this sense is not explicitly known. We thus study the original construction as a model for the confinement/deconfinement transition.} with Euclidean metric given by
\bea\label{wittenT}
&& ds^2 = \left(\frac{u}{R}\right)^{3/2}\left[{\tilde f}(u) dx_0^2 + dx_a dx^a + dx_4^2\right] + \left(\frac{u}{R}\right)^{-3/2}\left[\frac{du^2}{{\tilde f}(u)} + u^2 d\Omega_4^2\right]\,,\nonumber \\
&& {\tilde f}(u) = 1- \frac{u_T^3}{u^3}\,.
\eea
Here $a=1,2,3$. The $(x_0, u)$ subspace is a cigar - with the $S_{x_0}$ circle smoothly shrinking to zero size at $u=u_T$ (the position of the horizon) - provided we identify
\be
9 \beta^2 u_T = 16\pi^2 R^3\,.
\label{temp0}
\ee
The $(x_4, u)$ space is instead a cylinder. 
Notice that the two allowed background metrics are related by an exchange of $S_{x_0}$ with $S_{x_4}$ and of $u_T\sim T^2$ with $u_0\sim M_{KK}^2$ (notice in turn that these relations suggest a radius/energy relation of the form $u\sim E^2$). 
 
The background which minimizes the Euclidean on-shell gravity action $S$ will provide the dominant contribution to the gravity partition function in the classical limit $Z\sim e^{-S}$. 
It turns out that the black hole solution is the energetically preferred one when $T> T_c$, where the critical temperature $T_{c}$ is given by
\be
T_c = 1/\beta_4 = \frac{M_{KK}}{2\pi}\,.
\ee
At lower temperatures the other solution turns out to be energetically preferred. 

Each background solution is dual to a different phase of the corresponding field theory. 
At $T > T_c$ the theory turns out to be in a deconfined regime, while at $T<T_c$ the theory is confining. This can be explicitly checked by e.g. a holographic computation of the Polyakov loop. 

The holographic correspondence maps the difference between the on-shell Euclidean gravity actions on the two backgrounds with the free energy difference, among the two phases, in the dual large $N_c$, large $\lambda_4$ quantum field theory: $\Delta S = \beta \Delta F$. 
One can thus compute $\Delta F$ and study the phase diagram in a relatively simple way. In particular one can immediately verify that there is a first order phase transition at $T=T_c$.

When the $D8-{\bar D}8$-branes are placed at antipodal points on $S_{x_4}$ (the configuration we consider in this work), they fall into the horizon of the black hole in the $T>T_c$ phase, see figure \ref{fig0}.
Thus, they do not reconnect anymore and the flavor symmetry group remains $U(N_f)\times U(N_f)$: chiral symmetry is restored in the deconfined case and the transition temperature $T_c$ coincides with the confinement/deconfinement one \cite{Aharony:2006da}. 
When the $D8-{\bar D}8$-branes are not antipodal, deconfinement and chiral symmetry restoration generically happen at different critical temperatures \cite{Aharony:2006da}. 
Though this is an interesting possibility to consider, we leave the analysis of the corresponding setup for the future.

%%%%%%%%%%%%%%%%%%%%%%%%%%%%%%%%%%%%%%%%%%%
\section{The backreacted solution in the confined phase}\label{secconfinedcase}
\setcounter{equation}{0}
In \cite{ss} the flavor $D8$-branes were treated in the probe approximation. Here we want to take into account their backreaction on the background to first order in $N_f/N_c$. 
The trick we use in order to avoid having to deal with partial differential equations coupled to delta-function sources, it to consider a setup where a large number $N_f$ of $D8$-branes are homogeneously smeared along the transverse $x_4$ circle \cite{Bigazzi:2005md,Casero:2006pt,Nunez:2010sf,Bigazzi:2011db}. 
In this section we consider the whole model at $T < T_c$ and at finite quark chemical potential for the flavor fields. 

In our analysis (both in the confined and in the deconfined case) we will keep the length of the $x_4$ and $x_0$ circles fixed. 
The 't Hooft coupling $\lambda_4$, ``geometrically'' defined from the reduction of the 5d UV completion of the model on $x_4$, will thus be held fixed too.

\subsubsection*{Action and ansatz}
Our string frame metric ansatz for the $T=0$ case is the same one used in \cite{Aharony:2006da}\footnote{The relation with the coordinates used in (\ref{witten0}) is given in (\ref{errerho}), (\ref{solwitten})-(\ref{erreu}).}
\be
ds^2 = e^{2\lambda} (-dt^2 + dx_{a}dx^{a}) + e^{2\tilde\lambda}dx_4^2 + l_s^2 e^{-2\varphi}d\rho^2+l_s^2 e^{2\nu}d\Omega_4^2\,,
\label{anscon}
\ee
where $\lambda,\tilde\lambda,\varphi,\nu$ are functions of the radial coordinate $\rho$ and the $x_4$-coordinate is compactified on a circle of length
\be
\beta_4 = \frac{2\pi}{M_{KK}}\,.
\ee
When $T\neq 0$ what we need to do is just to compactify the Euclidean time on a circle of length $\beta=1/T$. Finally, the function $\varphi$ is related to the dilaton $\phi$ through the defining equation
\be
\varphi = 2\phi - 4\lambda - \tilde\lambda - 4\nu\,.
\ee
The relevant action from which the equations of motion follow is\footnote{Notice that in the following, since the $g_s$ factor is already contained in the on-shell value of $e^{\phi}$ as in the original Witten-Sakai-Sugimoto model, the $D8$-brane tension $T_8$ will be defined as $T_8=(2\pi)^{-8}\alpha'^{-9/2}$.}
\be\label{action0}
2k_0^2 S = \int d^{10}x\sqrt{-g}\left[e^{-2\phi}\left(R+4(\partial\phi)^2\right) -\frac12 |F_4|^2\right]-\frac{2k_0^2 N_f T_8 M_{KK}}{\pi}\int d^{10}x\frac{\sqrt{-g}}{\sqrt{g_{44}}} e^{-\phi}\,.
\ee
The last part of this action arises as the DBI contribution of $N_f$ U-shaped $D8$-branes which are homogeneously smeared on the transverse $x_4$ circle (whose metric component is $g_{44}$ and whose length is $2\pi/M_{KK}$) and reach the tip of the cigar.\footnote{Notice that we smear the U-shaped configuration along of the whole $x_4$ circle.} 
At leading order in $N_f/N_c$ further contributions from the bulk gravity fields will not enter the analysis and will hence be neglected (they are zero on-shell on the ansatz we use).\footnote{See e.g. \cite{Bigazzi:2011it,Cotrone:2012um,Bigazzi:2013jqa} for the similar situation in the $D3-D7$ system.} 

The smeared DBI action is put on-shell w.r.t. the embedding coordinate $x_4=x_4(\rho)$, which, in the confining case and for the antipodal configuration we are considering, satisfies the equation of motion $\dot x_4 =0$.
In order to account for the presence of  two branches at two antipodal points on the $x_4$ circle, we count the integration over the radial coordinate two times. 
Moreover, we put on-shell also the $U(1)$ gauge field on the branes, which is the holographic dual of the $U(1)_B$ current. A non trivial value of the electric component of the $U(1)$ field strength $F$ is related to a finite baryon density configuration. In the present case $F\equiv dA=0$: this in fact the relevant solution in the confining phase \cite{Kim:2006gp} at small baryon chemical potential $\mu$ and if no explicit sources are introduced.\footnote{The solution corresponds to a constant temporal component of the brane gauge field, $A_t = \mu$, which decouples from the remaining equations of motion. Notice that a constant value for the temporal component of the gauge field is allowed since, in the confining phase, the temporal circle does not shrink.}  
This corresponds to field theory configurations, which are encountered also in the QCD phase diagram, with finite quark chemical potential and zero baryon density.

In the probe approximation, with localized flavor branes, there are other possible relevant configurations of the gauge field.
For example, in the deconfined case the preferred configuration above a critical value of the charge density is spatially modulated \cite{Ooguri:2010xs}.
In the confined phase one can consider non-Abelian configurations and add to the setup $D4$-branes wrapped on $S^4$ which are instantons on the $D8$ worldvolume and act as sources (baryon vertices) of the $U(1)_B$ field (see for example \cite{bergman}). 
We will not consider these involved configurations in our context.

Implementation of the ansatz above (plus the usual ansatz in (\ref{witten0}) for $F_4$) gives the following 1d action
\bea
S &=& {\cal V}\int d\rho \left[-4{\dot\lambda}^2 - {\dot{\tilde\lambda}}^2 - 4{\dot\nu}^2 + {\dot\varphi}^2 + V\right]\,,\nonumber \\
V&=& 12 e^{-2\nu-2\varphi}-Q_c^2 e^{4\lambda+\tilde\lambda-4\nu-\varphi}-Q_f e^{2\lambda-\frac{\tilde\lambda}{2}+2\nu-\frac{3}{2}\varphi}\,,
\eea
which has to be supported by the zero-energy constraint
\be
-4{\dot\lambda}^2 - {\dot{\tilde\lambda}}^2 - 4{\dot\nu}^2 + {\dot\varphi}^2 - V=0\,.
\label{ze}
\ee
The equations of motion derived from the effective 1d action supported by this constraint coincide with the equations deduced from the 10d action (\ref{action0}) once the homogeneous ansatz (\ref{anscon}) is adopted. 

 Above we have defined (using $R^3 = \pi g_s N_c l_s^3$)
\be
Q_c = \frac{3}{\sqrt{2}g_s}\frac{R^3}{l_s^3}= \frac{3\pi N_c}{\sqrt{2}}\,,
\ee
as the constant arising from the quantized $F_4$ flux through $S^4$ and
\be
Q_f = \frac{2k_0^2 N_f T_8 M_{KK}l_s^2}{\pi}\,,
\ee
as the one related to the number $N_f$ of flavor branes. Finally the overall volume factor reads
\be
{\cal V}= \frac{1}{2k_0^2} V_3 V_{S^4}\frac{1}{T}\frac{2\pi}{M_{KK}}l_s^3\,,
\ee
where $V_3$ is the infinite 3d-space volume and $V_{S^4}= 8\pi^2/3$.

\subsubsection*{Equations of motion}
The equations of motion following from the previous action, which we re-arrange so that the dilaton $\phi$ appears instead of $\varphi$, are
\bea
&&\ddot{\lambda} -\frac{Q_c^2}{2}e^{8\lambda+2\tilde\lambda-2\phi}=\frac{Q_f}{4}e^{8\lambda+\tilde\lambda-3\phi+8\nu}\,,\nonumber \\
&&\ddot{\tilde\lambda} -\frac{Q_c^2}{2}e^{8\lambda+2\tilde\lambda-2\phi}=- \frac{Q_f}{4}e^{8\lambda+\tilde\lambda-3\phi+8\nu}\,, \nonumber \\
&&\ddot{\phi} -\frac{Q_c^2}{2}e^{8\lambda+2\tilde\lambda-2\phi}= \frac{5 Q_f}{4}e^{8\lambda+\tilde\lambda-3\phi+8\nu}\,, \nonumber \\
&&\ddot{\nu} +\frac{Q_c^2}{2}e^{8\lambda+2\tilde\lambda-2\phi}-3e^{8\lambda+2\tilde\lambda-4\phi+6\nu}= \frac{Q_f}{4}e^{8\lambda+\tilde\lambda-3\phi+8\nu}\,.
\eea
Notice that these imply that
\be
\ddot\chi=0\,\quad{\rm where}\,\,\,\chi\equiv 3\lambda-2\tilde\lambda-\phi\,.
\label{chieq}
\ee
In what follows it will be useful to define a new radial coordinate
\be\label{errerho}
r \equiv a\rho\,,\qquad a\equiv \frac{\sqrt{2} Q_c u_0^3}{3R^3 g_s}= \frac{u_0^3}{l_s^3 g_s^2}\,,
\ee
where $u_0$ is the minimal value of the original Witten-Sakai-Sugimoto radial variable $u$, i.e. the position of the tip of the $(x_4,u)$ cigar in the unflavored case. 

We are going to look for a perturbative solution of the above equations at first order in the effective parameter which weighs the flavor contribution to the action. In particular we will expand all the functions in the form
\be
\Psi(r) = \Psi_0(r) + \epsilon_f \Psi_1(r) + {\cal O}(\epsilon_f^2)\,,\
\ee
where
\be
\epsilon_f \equiv \frac{R^{3/2} u_0^{1/2} g_s}{l_s^2}Q_f = \frac{1}{12\pi^3}\lambda_4^2\frac{N_f}{N_c}\ll 1\,,
\label{ef}
\ee
is our expansion parameter and $\lambda_4 = g_{YM}^2 N_c =  2\pi g_s N_c l_s M_{KK}$ is the 't Hooft coupling at the UV scale set by $M_{KK}$.
The zero-th order unflavored solutions in these coordinates can be read from \cite{Aharony:2006da} and are given by
\bea\label{solwitten}
&&\lambda_0 (r) = f_0 (r) + \frac{3}{4} \log{\frac{u_0}{R}}\,,\nonumber \\
&& \tilde\lambda_0(r) = f_0(r) -\frac{3}{2} r +  \frac{3}{4} \log{\frac{u_0}{R}}\,,\nonumber \\
&& \phi_0 (r) = f_0(r) + \frac{3}{4} \log{\frac{u_0}{R}}+\log g_s\,,\nonumber \\
&& \nu_0(r) = \frac{1}{3}f_0(r) + \frac{1}{4} \log{\frac{u_0}{R}}+\log\frac{R}{l_s}\,,
\eea
with
\be
f_0(r) = -\frac{1}{4}\log\left[1-e^{-3r}\right]\,.
\ee
Notice that for the combination $\chi$ defined in (\ref{chieq}) we have
\be
\chi_0 \equiv 3\lambda_0-2\tilde\lambda_0-\phi_0 = -\log g_s + 3r\,.
\label{chi0}
\ee
In the unflavored case, the variable $r$ is related to the standard Witten-Sakai-Sugimoto radial variable by 
\be\label{erreu}
e^{-3r} = 1-\frac{u_0^3}{u^3}\,,
\ee
so that $r\rightarrow 0$ (resp. $r\rightarrow\infty$) when $u\rightarrow\infty$ (resp. $u\rightarrow u_0$).

The equations of motion for the first order terms read (derivatives are with respect to $r$ now)
\bea
&&\lambda_1'' - \frac{9}{2}\frac{e^{-3r}}{(1-e^{-3r})^2} (4\lambda_1 +\tilde\lambda_1-\phi_1) = \frac{1}{4}\frac{e^{-\frac32 r}}{(1-e^{-3r})^{13/6}}\,,\nonumber \\
&&\tilde\lambda_1'' - \frac{9}{2}\frac{e^{-3r}}{(1-e^{-3r})^2} (4\lambda_1 +\tilde\lambda_1-\phi_1) = -\frac{1}{4}\frac{e^{-\frac32 r}}{(1-e^{-3r})^{13/6}}\,,\nonumber \\
&&\phi_1'' - \frac{9}{2}\frac{e^{-3r}}{(1-e^{-3r})^2} (4\lambda_1 +\tilde\lambda_1-\phi_1) = \frac{5}{4}\frac{e^{-\frac32 r}}{(1-e^{-3r})^{13/6}}\,,\nonumber \\
&&\nu_1'' - \frac{3}{2}\frac{e^{-3r}}{(1-e^{-3r})^2} (4\lambda_1 +\tilde\lambda_1-5\phi_1+12\nu_1) = \frac{1}{4}\frac{e^{-\frac32 r}}{(1-e^{-3r})^{13/6}}\,.
\eea
From these we see that
\bea\label{lambdaphi}
&&\tilde\lambda_1 = \lambda_1-\frac{1}{2} f - A_1 - B_1\, r\,,\nonumber \\
&& \phi_1 = \lambda_1+ f - A_2 - B_2\, r\,,
\eea
where $f(r)$
%\be
%f(r) = \frac{4}{9} e^{-3r/2} \, _3F_2\left(\frac{1}{2},\frac{1}{2},\frac{13}{6};\frac{3}{2},\frac{3}{2};e^{-3r}\right)\,,
%\ee
is a particular solution 
%(in terms of a generalized hypergeometric function) 
of
\be\label{effedue}
f '' (r) = \frac{e^{-3r/2}}{(1- e^{-3r})^{13/6}}\,,
\ee
and $A_{1,2}, B_{1,2}$ are integration constants. 

\subsubsection*{Solution and asymptotics}

Plugging the expressions (\ref{lambdaphi}, \ref{effedue}) in the remaining equations and integrating, we find the solution
\bea
&&\lambda_1 = \frac38 f + y -\frac14(A_2-A_1)-\frac14(B_2-B_1) r\,\nonumber \\
&&\tilde \lambda_1 = -\frac18 f + y -\frac14(A_2+B_2 r)-\frac34(A_1+B_1 r) \, \nonumber\\
&&\phi_1 = \frac{11}{8} f + y +\frac14(A_1+B_1 r)-\frac54(A_2+B_2 r) \,\nonumber \\
&& \nu_1 = \frac{11}{24}f + q\,,
\eea
with
\bea \label{effe}
&& f = \frac{4}{9} e^{-3r/2} \, _3F_2\left(\frac{1}{2},\frac{1}{2},\frac{13}{6};\frac{3}{2},\frac{3}{2};e^{-3r}\right)\,,\nonumber \\
&& y = C_2 -\coth \left(\frac{3r}{2}\right) \left(C_1+C_2 \left(\frac{3r}{2}+1\right)\right) + z\,,\nonumber \\
&& q =  \frac{1}{12} (A_1-5 A_2+r (B_1-5 B_2))+\frac{5}{3}z- \coth \left(\frac{3
   r}{2}\right) (M_1+M_2 (3 r+2))+2 M_2\,,\nonumber \\
&& z  =  -\frac{e^{-9 r/2} \left(e^{-3 r}+1\right) \left(9 e^{3 r} \,
   _3F_2\left(\frac{1}{2},\frac{1}{2},\frac{19}{6};\frac{3}{2},\frac{3}{2};e^{-3
   r}\right)+\,
   _3F_2\left(\frac{3}{2},\frac{3}{2},\frac{19}{6};\frac{5}{2},\frac{5}{2};e^{-3
   r}\right)\right)}{162 \left(1-e^{-3 r}\right)} \nonumber\\
   && \qquad -\frac{8 e^{-3 r/2} \left(10 e^{-3 r}+3\right)
   \, _2F_1\left(\frac{1}{6},\frac{1}{2};\frac{3}{2};e^{-3 r}\right)}{819
   \left(1-e^{-3 r}\right)}+ \frac{e^{-15 r/2} \left(38 e^{3 r}+8 e^{6
   r}-40\right)}{273 \left(1-e^{-3 r}\right)^{13/6}}\,,
\eea
given in terms of (generalized) hypergeometric functions. All in all we have eight integration constants $A_{1,2},B_{1,2},C_{1,2},M_{1,2}$. Some of them are fixed by physical requirements.

The zero energy constraint (\ref{ze}) is satisfied, to first order in $\epsilon_f$, provided the condition
\be
5B_1-B_2-18(C_2+4M_2)=0
\label{cze}
\ee
holds.

Other constraints arise by requiring regularity at the tip of the $(x_4,r)$ cigar, which corresponds to the $r\rightarrow\infty$ limit. In this case the corresponding (IR) behavior of the various functions above can be easily extracted working with the variable $x=e^{-3r/2}$, which goes to zero in the limit. As a result we find the following IR asymptotics
\bea
&&\lambda_1 = \frac{1}{12} \left[3 (A_1 - A_2 - 4 C_1) + 2 (-B_1 + B_2 + 6 C_2) \log x \right] + {\cal O}(x^2)\,,\nonumber \\
&&\tilde\lambda_1 = \frac{1}{12} \left[-3 (3 A_1 + A_2 + 4 C_1) + 2 (3 B_1 + B_2 + 6 C_2) \log x\right] + {\cal O}(x^2)\,,\nonumber \\
&&\phi_1 = \frac{1}{4} (A_1 - 5 A_2 - 4 C_1) +\frac{1}{6} (-B_1 + 5 B_2 +6 C_2) \log x +{\cal O}(x^2)\,,\nonumber \\
&&\nu_1 = \frac{1}{36} \left[3 (A_1 - 5 A_2 - 12 M_1) - 2 (B_1 - 5 B_2 - 36 M_2) \log x\right] + {\cal O}(x^2)\,.
\eea
Regularity at the tip of the cigar can be achieved imposing the following conditions
\be
B_1=6C_2\,,\quad B_2=0\,,\quad M_2=\frac{C_2}{6}\,,
\label{IRreg}
\ee
which could be further restricted by requiring
\be
C_2=0\,,
\label{eqc2}
\ee
in order to drop all the IR logarithmically divergent factors from the metric components and the dilaton; this is not strictly necessary: the logarithmically divergent factor in $\tilde\lambda_1$ is anyway subleading, in the small $\epsilon_f$ limit, w.r.t. the analogous term in $\tilde\lambda_0$. 
It is notable that if condition (\ref{IRreg}) is satisfied, the constraint (\ref{cze}) following from the zero energy condition is automatically satisfied too.

Further considerations on the integration constants can be done by observing that, for the IR regular solutions selected by (\ref{IRreg})
\be
\chi_1\equiv 3\lambda_1-2\tilde\lambda_1-\phi_1 = (2A_1+A_2) + 12C_2\, r\,,
\label{chi1}
\ee
consistently with the equation $\ddot\chi=0$ valid to all orders in $\epsilon_f$. Comparing (\ref{chi1}) with its zero-th order relative (\ref{chi0}) we see that $C_2$ can be interpreted as a rescaling of the radial coordinate, while $2A_1+A_2$ can be read as a correction to the string coupling $g_s$. With an an all-order choice of the two integration constants related with $\chi$, one could fix $\chi=\chi_0$, obtaining, for the case at hand, the conditions $C_2 =0$ and $2A_1+A_2=0$.

The UV ($r\rightarrow0$, i.e. $x\rightarrow1$) behavior of the above functions is given by
\bea \label{UVasympt}
&&\lambda_1 = \frac{-C_1- C_2+ k}{1-x}+\frac{101}{455 (2)^{1/6}(1-x)^{1/6}}+ \lambda_1^{(UV,0)}+ {\cal O}\left((1-x)^{5/6}\right)\,,\nonumber \\
&&\tilde\lambda_1= \frac{- C_1- C_2+k}{1-x}-\frac{29}{455
   (2)^{1/6}(1-x)^{1/6}}+ \tilde\lambda_1^{(UV,0)}+{\cal O}\left((1-x)^{5/6}\right)\,,\nonumber \\ 
&&\phi_1=\frac{- C_1- C_2+k}{1-x}+\frac{361}{455 (2)^{1/6}(1-x)^{1/6}}+ \phi_1^{(UV,0)}+ {\cal O}\left((1-x)^{5/6}\right)\,,\nonumber \\
&&\nu_1= \frac{\frac{5 k}{3}-M_1-2
   M_2}{1-x}+ \frac{25}{91 (2)^{1/6}
   (1-x)^{1/6}}+\nu_1^{(UV,0)}+ {\cal O}\left((1-x)^{5/6}\right)\,, 
\eea
where
\bea
&&\lambda_1^{(UV,0)}= \frac{1053 \Gamma \left(-\frac{2}{3}\right)^2
   (A_1- A_2+2 (C_1+ C_2))}{4212 \Gamma
   \left(-\frac{2}{3}\right)^2} +  \\ 
&& \qquad \qquad  + \frac{10 (2)^{1/3} \pi 
   \left[-681+85 \sqrt{3} \pi +255 \log \left(\frac{27}{16}\right)\right] \Gamma
   \left(-\frac{10}{3}\right)}{4212 \Gamma
   \left(-\frac{2}{3}\right)^2}  \,,\nonumber \\
&&\tilde\lambda_1^{(UV,0)}=  \frac{10 (2)^{1/3} \pi  \left[3 (85+76 \log (2)-57 \log (3))-19 \sqrt{3} \pi
   \right) \Gamma \left(-\frac{10}{3})\right]}{4212 \Gamma
   \left(-\frac{2}{3}\right)^2}+\nonumber \\
&&  \qquad \qquad -\frac{1053 \Gamma
   \left(-\frac{2}{3}\right)^2 (3 A_1+A_2-2
   (C_1+ C_2))}{4212 \Gamma
   \left(-\frac{2}{3}\right)^2}\,, \nonumber\\
&&\phi_1^{(UV,0)}= \frac{A_1}{4}-\frac{5
   A_2}{4}+\frac{C_1}{2}+\frac{C_2}{2}+\frac{5 \pi\left[-2553+293 \sqrt{3} \pi +879 \log \left(\frac{27}{16}\right)\right]
   \Gamma \left(-\frac{10}{3}\right)}{1053 (2)^{2/3} \Gamma\left(-\frac{2}{3}\right)^2}\,,\nonumber \\ 
&&\nu_1^{(UV,0)}=\frac{1}{12} (A_1-5 A_2+6 M_1)+M_2+ \frac{\sqrt{\pi } \left[-823+107 \sqrt{3} \pi +321 \log\left(\frac{27}{16}\right)\right] \Gamma \left(-\frac{7}{6}\right)}{18720\Gamma \left(-\frac{5}{3}\right)}\,, \nonumber
\eea        
and 
\be
k = \frac{\pi ^{3/2} \left(3+\sqrt{3} \pi -12 \log (2)+9 \log (3)\right)}{78 \Gamma
   \left(-\frac{2}{3}\right) \Gamma \left(\frac{1}{6}\right)}\,.
\ee   
It is interesting to notice that the subleading UV divergences, that is the terms diverging as $(1-x)^{-1/6}$, do not depend on the integration constants. 
We interpret these as the dual of the ``universal" terms related to the addition of the flavors to the original unflavored action. 
As in the $D3-D7$ case \cite{Benini:2006hh}, the flavor source term induces a Landau Pole in the UV. This is the reason why these subleading divergences cannot be killed by a choice of integration constants. 

The combinations of integration constants appearing in the UV asymptotics, can instead be interpreted as corresponding to sources or VEVs of other gauge invariant operators. The sources can be turned off by a suitable choice of the integration constants. We defer the field/operator analysis for the future, so we do not know yet what are the combinations of $\lambda_1, \tilde\lambda_1, \phi_1, \nu_1$ corresponding to gauge invariant operators.
Nevertheless, we are interested in switching off at least the most divergent terms in (\ref{UVasympt}), hence we impose the prudent condition 
\be
C_1+C_2 =k\,, \qquad M_1+2M_2 = \frac{5}{3}k \,.   
\label{UVsources}
\ee
A supplementary option is to make sure that all the subleading terms are turned off, which requires the condition
\be
\lambda_1^{(UV,0)}=\tilde\lambda_1^{(UV,0)}=\nu_1^{(UV,0)}=\phi_1^{(UV,0)}=0\,.
\label{UVvevs}
\ee
These conditions, together with (\ref{UVsources}), imply that
\be
A_1=\frac{81 \sqrt{3} \pi ^2 \left(-9+\sqrt{3} \pi -12 \log (2)+9 \log
   (3)\right)}{43120 (2)^{2/3} \Gamma \left(-\frac{14}{3}\right) \Gamma
   \left(-\frac{2}{3}\right)^2}\,,\quad A_2=-2A_1\,, 
\label{A1A2}
\ee
the latter being consistent with the choice $\chi=\chi_0$ on the integration constants. In the following we will consider the family of solutions for which the IR regularity conditions (\ref{IRreg}), and the conditions (\ref{UVsources}) killing the leading UV divergences hold. On this family we thus have
\be
B_1=6 C_2\,, \quad B_2=0\,, \quad M_2=\frac{C_2}{6}\,, \quad C_1=k-C_2\,,\quad M_1 =(5k-C_2)/3\,.
\ee
We will take $A_1,A_2, C_2$ generic, having in mind their special values (\ref{eqc2}), (\ref{A1A2}) along the analysis.
%%%%%%%%%%%%%%%%%%%%%%%%%%%%%%%%%%%%%
\subsection{Some physical properties}\label{secphys}

In this section we give a flavor of the physical effects which can be studied with the backreacted solution.

Let us begin by analyzing how the flavor contributions modify the relation between the radial parameter $u_0$ and the mass $M_{KK}$. This relation comes from the requirement that the $(x_4,r)$ cigar closes smoothly at the tip (which is at $r\rightarrow\infty$) and reads
\be
\frac{u_0}{R^3} = \frac49 M_{KK}^2 \left[1+\frac{\epsilon_f}{3}(5A_1-A_2 -4 C_2 -52 k)\right]\,.
\label{u0Mkk}
\ee
For the special choices (\ref{eqc2}), (\ref{A1A2}), $5A_1-A_2 -4C_2-52k\approx 1.99$. 

\subsubsection*{The running coupling and the UV Landau Pole}
The running gauge coupling can be obtained by examining the action of a probe $D4$-brane wrapped on the $x_4$ circle (see e.g. \cite{noiWL}). It reads (using the $x=e^{-3r/2}$ variable)
\be
\frac{1}{g^2_{YM,\,x}} =\frac{1}{ 2\pi l_s M_{KK,0}} e^{-\phi+ \tilde\lambda} = \frac{x}{g_{YM}^2}\left[1-\epsilon_f(\phi_1-\tilde\lambda_1)\right]\,.
\ee
In the UV limit $x\rightarrow 1$
\be
 \frac{1}{g^2_{YM,\,x}} \approx \frac{1}{g_{YM}^2} \left[ 1-\frac37\epsilon_f \frac{2^{5/6}}{(1-x)^{1/6}} \right]\,,
 \ee
hence, differently from the unflavored case (where it goes to the constant ``geometrical" value
$g_{YM}^2$), the running coupling tends to diverge in the UV, signaling the presence of a Landau pole. The gauge theory analysis can thus be safely performed only in the IR, that is at scales
\be\label{LPposition}
x\ll x_{LP}\,,\qquad {\rm where}\qquad x_{LP} = 1 - 2^5 (3/7)^6 \epsilon_f^6 \,.
\ee
Note that in the perturbative analysis we are considering (where $\epsilon_f\ll1$) the UV Landau Pole essentially coincides with the asymptotic value $x\rightarrow1$.
\subsubsection*{The string tension}
Using (\ref{u0Mkk}), the string tension turns out to be given by\footnote{See also \cite{giata} for other studies on flavor corrections to the static potential in the WSS model.}
\be
T_s = \frac{1}{2\pi\alpha'} e^{2\lambda}|_{x=0}=\frac{2}{27\pi}\lambda_4 M_{KK}^2 \left[1+\epsilon_f(3A_1-A_2-28k)\right]\,.
\ee
Notice that this does not depend on $C_2$. For the special choice of integration constants (\ref{A1A2}), $3A1-A2-28k\approx 1.13$. 
In this case one could conclude, naively, that the effect of the dynamical flavors is that of increasing the string tension. 
However, as we will also discuss in the following, care as to be taken when comparing theories with different number of flavors.  

\subsubsection*{Baryon mass}
In the model at hand, a baryon vertex is identified \cite{Witten:1998xy} with a Euclidean $D4$-brane wrapped on $S^4$ and localized at the radial position corresponding to the deep IR of the dual field theory. 
Using the backreacted metric found above we can easily study how the dynamical flavors affect the mass of the baryon. The wrapped $D4$-brane action reads
\be
S^{(E)}_{D_4}= T_4 \int dx^0 d\Omega_4 e^{-\phi}\sqrt{\det g_5} = T_4 V(S^4) l_s^4 \int dx^0 e^{\lambda+4\nu-\phi}|_{x=0}\equiv m_B\int dx^0\,,
\ee 
where $T_4=(2\pi)^{-4}l_s^{-5}$ is the $D4$-brane tension and $x=e^{-3r/2}\rightarrow0$
is the IR value of the radial variable. Using our solution we thus get that the baryon mass is given by
\be
m_B = \frac{1}{27\pi}\lambda_4 N_c M_{KK}\left[1+\epsilon_f\left(2A_1 - A_2 - 24k\right)\right]\,.
\ee
For the special choices (\ref{A1A2}), $2A_1-A_2-24k \approx 0.95$.

\subsubsection*{Vector meson mass spectrum}
Following the same reasonings as in Section 3.3 of \cite{ss} (to which we refer the interested reader for details), the spectrum of the massive vector mesons on the backreacted background can be found by considering the fluctuation of a gauge field on a probe $D8$-brane. The massive vector mesons have masses given by
\be
m_n^2 = \frac94\frac{u_0}{R^3}\gamma_n = M_{KK}^2 \left[1+\frac{\epsilon_f}{3}(5A_1-A_2 -4C_2 -52 k)\right]\gamma_n\,,\qquad n\ge1\,,
\ee
where $\gamma_n$ are given by the equation
\be
\partial_r \left[ F(r) \partial_r \psi_n\right] + H(r) \gamma_n \psi_n = 0\,,
\label{eqmes}
\ee
where
\bea\label{FH}
&&F(r)= e^{3r/2}\sqrt{1-e^{-3r}}\left[1+\epsilon_f (\phi_1-2\lambda_1-\tilde\lambda_1)\right]\,,\nonumber \\
&& H(r) = \frac{9}{4}\frac{e^{-3r/2}}{(1-e^{-3r})^{7/6}}\left[1+\epsilon_f (-3\phi_1+4\lambda_1+\tilde\lambda_1+8\nu_1)\right]\,,
\eea 
and the functions $\psi_n$ satisfy the normalization condition
\be
A\int dr H(r) \psi_n \psi_m = \delta_{n\,m}\,,
\ee
with
\be
A = \frac{T_{8} V_{S^4} (2\pi)^2 l_s^4 R^{9/2} u_0^{1/2}}{g_s}\,.
\ee
Normalizability forces us to choose an asymptotic condition such that the modes $\psi_n$ vanish in the UV. 
As we have already noticed, the flavored theory we are considering has a Landau Pole in the UV and hence the same condition has to be imposed at a UV cutoff.
In the deep IR (i.e. at $x=0$ or $r\rightarrow\infty$) we impose the regularity conditions $\psi=0$ or $\psi '=0$ which distinguish among even and odd functions. 
Correspondingly we have vector and axial-vector mesons. 
Modes with $n$ odd (resp. even) correspond to the former (resp. the latter). 
As in \cite{ss} we have that the lightest mode $\gamma_1$ corresponds to a vector meson (the ``$\rho$'') with $C=P=-1$, the second one $\gamma_2$ corresponds to an axial-vector meson with $C=P=+1$ (the ``$a_1(1260)$'') and so on. 

We solve equation (\ref{eqmes}) with the standard shooting technique.
We do not perform a scan of the results as the parameters of the solution are varied, but consider one relevant case.
Namely, the integration constants are fixed such that: i) all the IR logarithmical divergences are canceled (i.e. we enforce the special condition $C_2=0$); ii) the free energy coincides with the one obtained in the probe calculation (as expected, see the next section, formula (\ref{thermoconsist})); iii) the source for the operator dual to the field $\lambda_1-\tilde\lambda_1$ is turned off.
The latter combination enters naturally the reduction of the ten dimensional action to five dimensions (where the unflavored metric is asymptotically $AdS$ in the dual frame \cite{skenderis}), giving a well defined scalar in the reduced theory, so it is very likely to have a dual operator without further mixing with other modes. 
Its source depends only on the constant $A_1$.
Conditions ii) and iii) turn out to coincide with (\ref{A1A2}), giving $A_1\sim 0.047, A_2 \sim -0.094$.

The results for the vector meson spectra are as follows.
The ratio of the masses of the first two $\rho$ mesons is reduced by the flavor contribution,\footnote{The same reduction was observed on the flavored version of other theories in \cite{Bigazzi:2009gu}. There, it was also noted that this effect is true only in the naive comparison scheme, which we are tacitly using in this section, where no scale or observable is kept fixed when comparing the unflavored and flavored theories. The effect can be qualitatively different by using a different comparison scheme, see section \ref{sectemperature}.} giving a result closer to the phenomenological one than the unflavored theory.
Namely, while experimentally $m^2_{\rho(1450)}/m^2_{\rho} \sim 3.5$ and in the Witten-Sakai-Sugimoto model this ratio is around $4.3$, in the flavored case with $\epsilon_f =0.02$ (taken as a representative value) it is $3.7$.
Also the ratio of the masses of the first axial vector with the $\rho$ meson is reduced by the flavor contribution, giving a result which is now more distant from the phenomenological one than the unflavored theory.
That is, from experiments $m^2_{a_1(1260)}/m^2_{\rho} \sim 2.51$, in the Witten-Sakai-Sugimoto model this value is $2.39$ and in the flavored case with $\epsilon_f =0.02$ it is $2.37$.
As in the unflavored theory, the masses of higher mesons are quite larger than the experimental ones.

In the unflavored case, it is possible to look for approximate solutions of the equations (\ref{eqmes}) also using the WKB method as reported e.g. in \cite{russo}. The result should hold in principle only for higher modes, but it is ``massaged" in order to account for the small $n$ behavior too. It reads
\be
\gamma_n^{WKB,0}= \frac{\pi^2}{\xi_0^2}\left(\frac{n+1}{2}\right)^2\,,
\ee
where
\be
\xi_0=\int dr \sqrt{\frac{H_0(r)}{F_0(r)}} = \frac{3\sqrt{\pi}\Gamma[7/6]}{\Gamma[2/3]}\,,
\ee
with $H_0, F_0$ being the functions defined in (\ref{FH}) in the $\epsilon_f=0$ case. The WKB approximation hence gives
\be
\gamma_{1,2,3,4}^{WKB,0}\approx 0.74\,, 1.67\,,2.97\,,4.65\,,
\ee
to be compared with the numerical results $0.67\,, 1.6\,, 2.9\,, 4.5$ obtained in \cite{ss}. 

In the flavored case the WKB analysis is harder. One can imagine that (especially for low lying modes) the meson mass ratios do not depend on the $\epsilon_f$ parameter (as the $\epsilon_f$ dependence should be encoded in the overall scale). If this is the case one could guess that
\be
\gamma_n^{WKB} = \frac{\pi^2}{\xi^2}\left(\frac{n+1}{2}\right)^2\,,
\ee
where $\xi$ is the flavored version of $\xi_0$
\be
\xi=\int dr \sqrt{\frac{H(r)}{F(r)}}\,.
\ee
Setting $C_2=0$ we find
\be
\xi=\xi_0\left[1-\frac{\epsilon_f}{6}(A_1+A_2)\right] - \epsilon_f {\cal I}\,,
\ee
where ${\cal I}$ is a divergent integral. This indicates that a careful treatment of the solution is needed by introducing a cutoff and expanding the solution below this cutoff.
We hope to come back to this issue in the future.

%%%%%%%%%%%%%%%%%%%%%%%%%%%%%%%%%%
\section{Holographic renormalization of the Witten-Sakai-\\ Sugimoto model}
\label{holren}

In this section we present the holographic renormalization of the Witten-Sakai-Sugimoto model, both in the probe approximation and in the backreacted case.
As far as we know, the covariant counterterms needed to renormalize the action of this very well known model are not present in the literature. 
In the paper \cite{ballon}, where the WSS model in an external magnetic field is considered, counterterms for the probe $D8$-branes are provided, but they are not written in a covariant way.

A contingent reason why we need to holographically renormalize the theory is that we want to compute the free energy of the model. 
This is related to the renormalized on-shell Euclidean gravity action through the holographic formula $F = T S_{E,on-shell}^{ren}$.  Here
\be
S_{E}^{ren} = (S_{E} + {S}_{GH})+S_{c.t.}^{bulk}+S_{c.t.}^{D8}\,,
\ee
where $S_{E}$ is the Euclidean version of the bulk+smeared-$D8$-brane action
\be \label{totalaction}
S_{E}= -\frac{1}{2k_0^2}\left[\int d^{10}x\sqrt{g}\left[e^{-2\phi}\left(R+4(\partial\phi)^2\right) -\frac12 |F_4|^2\right]-\frac{Q_f}{l_s^2}\int d^{10}x\frac{\sqrt{g}}{\sqrt{g_{44}}} e^{-\phi}\right]\,.
\ee
Notice that we work in string frame.
The standard Gibbons-Hawking term reads\footnote{As a correction to formula (E.11) in \cite{myersmateos} we have verified that the terms in the gradient of the dilaton sum up to give a zero contribution.}
\be \label{gibbonshawking}
S_{GH} =   -\frac{1}{k_0^2}\int d^{9}x\sqrt{h}e^{-2\phi}K\,,
\ee
%2n\cdot \nabla\phi
where $h$ is the determinant of the boundary metric (in our coordinates it is the slice of the 10d metric at fixed $r=\epsilon$ with $\epsilon\rightarrow0$ being the UV boundary), $K$ is the trace of the extrinsic curvature of the boundary, which in our case is explicitly given by
\be
K = h^{MN}\nabla_M n_N = - \frac{1}{\sqrt{g}}\partial_r\left(\frac{\sqrt{g}}{\sqrt{g_{rr}}}\right)|_{r =\epsilon}\,,
\ee
and 
\be
n^{M} = - \frac{\delta^{M\,r}}{\sqrt{g_{rr}}}
\ee
is an outward pointing unit normal vector to the boundary at $r=\epsilon$ (the minus sign originates from the fact that our coordinate $r$ decreases towards zero when approaching the boundary).
The ``bulk" counterterm is a volume boundary term, already introduced in \cite{myersmateos}
\be \label{bulkcounterterm}
S^{bulk}_{c.t.}=\frac{1}{k_0^2}\left(\frac{g_s^{1/3}}{R}\right)\int d^9x \sqrt{h}\,\frac52 e^{-7\phi/3}\,.
\ee
Evaluating the above terms on our solution we get
\bea
&& S_{E}= - a{\cal V}\left[\frac94 -\frac{3}{2\epsilon}+\epsilon_f h_1\right]\,,\\
&& S_{GH} =  - a{\cal V}\left[-\frac74 +\frac{19}{6\epsilon}+\epsilon_f h_2\right]\,,\\ 
&& S^{bulk}_{c.t.} =  a{\cal V}\left[\frac{5}{3\epsilon}+\epsilon_f h_3\right]\,, 
\eea
where
\bea
&& a{\cal V}=\frac{1}{2k_0^2g_s^2}\frac{V_3}{T}\frac{2\pi}{M_{KK}}V_{S^4}u_0^3\,, \\
&& h_1 = 9 C_2-\frac{150\pi^{3/2}}{7\Gamma\left(-\frac23\right)\Gamma\left(\frac16\right)} -\frac{823}{1365 (3)^{1/6}\epsilon^{7/6}}-\frac{955}{132(3)^{1/6}\epsilon^{1/6}}\,, \nonumber \\
&& h_2 = -7 C_2 +\frac{22\pi^{3/2}}{\Gamma\left(-\frac23\right)\Gamma\left(\frac16\right)}+\frac{25}{39 (3)^{1/6}\epsilon^{7/6}}+\frac{959}{132(3)^{1/6}\epsilon^{1/6}}\,,\nonumber \\
&& h_3 = \frac{14}{117 (3)^{1/6}\epsilon^{7/6}}+\frac{35}{198(3)^{1/6}\epsilon^{1/6}}\,.
\eea
In the unflavored case, the bulk counterterm precisely cancels the $1/\epsilon$ divergence from $S_E+S_{GH}$ so that
\be
S_{E}^{ren}|_{N_f=0} = - \frac12 a{\cal V}\,,
\label{ren0}
\ee
consistently with equation (E.13) in \cite{myersmateos}. Correspondingly the free energy density reads (using (\ref{uomp}))\footnote{We are confident that the notation ``$f$'' for the free energy density will not be confused with the function $f$ in the solution (\ref{effe}) because of the context.}
\be
f_{N_f=0} = -\frac{2 N_c^2 \lambda_4}{3^7 \pi^2}M^4_{KK}\,.
\ee
The novelty in our analysis at this point is that in the flavored case there are (apart from novel constant terms) two classes of novel divergences (in $\epsilon^{-1/6}$ and $\epsilon^{-7/6}$) and the counterterm we have introduced above is not enough to cancel them
\be
h_1+h_2-h_3 = 2 C_2 +\frac47\frac{\pi^{3/2}}{\Gamma\left(-\frac23\right)\Gamma\left(\frac16\right)}-\frac{334}{4095 (3)^{1/6}\epsilon^{7/6}}-\frac{29}{198 (3)^{1/6}\epsilon^{1/6}}\,.
\ee
We thus need to introduce new counterterms $S_{c.t.}^{D8}$  for the $D8$-brane contribution.

%%%%%%%%%%%%%%%%%%%%%%%%%%%%%%%%%%%%%%%%%%%%%
\subsection{The probe approximation}
In the probe approximation, the bulk action is renormalized through the addition of the boundary terms introduced above, giving rise to (\ref{ren0}) as a result. 
To this we have to add the on-shell $D8$-brane action ($g_{9,MN}$ is the induced metric on the worldvolume; its on-shell value on the U-shaped embedding is just the $x_4=const.$ slice of the original background geometry)
\be
S_{D8} = \frac{Q_f}{2k_0^2}\frac{2\pi}{l_s^2 M_{KK}}\int \d^9x e^{-\phi_0}\sqrt{det g^{(0)}_9} = a\,{\cal V}\, \epsilon_f\, d_{1,probe} \, ,
\label{sd8con}
\ee
where 
\be
d_{1,probe}=\frac{2}{21(3)^{1/6}\epsilon^{7/6}}+\frac{7}{6(3)^{1/6}\epsilon^{1/6}}+\frac{24}{7}\frac{\pi^{3/2}}{\Gamma\left(-\frac23\right)\Gamma\left(\frac16\right)}
+{\cal O}(\epsilon)\,.
\label{dived8}
\ee
The on-shell DBI action thus has two divergent terms in the UV and needs to be renormalized.    

We get the required counterterms in the probe approximation, first going to the ``dual frame'' metric $d\tilde s^2 \sim e^{-2\phi/3}ds^2$ \cite{skenderis} and then reducing on $S^4$.\footnote{See also \cite{Benincasa:2009ze} for the application of the procedure for other probe branes.} 
In the dual frame, the metric is asymptotically $AdS$, so one can figure out the counterterms as the standard volume and GH ones on that background.
The resulting counterterm boundary action (already uplifted back to 8d)
\be
\tilde S_{c.t.}\sim \int d^8x e^{2\phi} \sqrt{{\tilde h}_8}\left[m_1- 2 m_2{\tilde K_{9}}\right]\,,
\ee
can be written in terms of covariant pieces in the induced boundary metric in the dual frame. Going back to the original metric and adapting everything to the smeared case, the resulting counterterms read
\be \label{D8couterterm}
S_{c.t.}^{D8} = \frac{Q_f}{k_0^2 l_s^2}\int d^9x \frac{\sqrt{h}}{\sqrt{h_{44}}}\left[ \frac{R}{g_s^{1/3}} m_1\, e^{-2\phi/3}- 2 m_2\frac{R^2}{g_s^{2/3}}\, e^{-\phi/3}\left( K  -\frac83 n\cdot\nabla\phi  - n\cdot\frac{\nabla(\sqrt{g_{44}})}{\sqrt{g_{44}}}\right)\right]\,.
\ee
Evaluating this explicitly one gets
\be
S_{c.t.}^{D8}=a{\cal V} \epsilon_f \left[\frac{2m_1- 8 m_2}{3(3)^{1/6}\epsilon^{7/6}}+\frac{14m_1- 8 m_2}{12(3)^{1/6}\epsilon^{1/6}}\right]\,.
\label{ctdp}
\ee
In order to cancel the divergences in the probe $D8$ action (\ref{dived8}) we have to choose%\footnote{A first computation using the reduced dual frame action, gave $m_1=-8/7$ and $m_2=-1/4$ in the confining phase we are focusing on and $m_1=1/7$, $m_2=1/14$ in the deconfined phase.}
\be
m_1 = - \frac{8}{7}\,,\quad m_2= -\frac14\,.
\label{m12p}
\ee
Factors in $g_s, l_s$ and $R$ appearing in the counterterms are remnants of the common asymptotic conditions (e.g. on the flux of $F_4$) for the backgrounds for which these counterterms have to be added. 

For the reader convenience, the counterterm for the $D8$-branes in standard eight dimensional form reads\footnote{Remember that in our notation $T_8=(2\pi)^{-8}\alpha'^{-9/2}$.}
\be 
S_{c.t.}^{D8} = - 2 N_f T_8 \int d^8x \sqrt{h_8}\left[ \frac{16}{7}\frac{R}{g_s^{1/3}}\, e^{-2\phi/3}-\frac{R^2}{g_s^{2/3}}\, e^{-\phi/3}\left( K_9  -\frac83 n\cdot\nabla\phi \right)\right]\,.
\ee
Let us outline that the $D8$-counterterms we are proposing here, although covariant and (automatically) local in the embedding field $x_4(r)$, contain radial derivatives of the background metric and dilaton on the boundary. It should be worth to reconsider these conterterms within an effective 1d Hamiltonian formulation, as in \cite{papad}, where they are expected to be expressed in terms of just the 1d boundary fields (and not of their conjugate momenta). We hope to come back to this issue in the future.\footnote{We are grateful to Ioannis Papadimitriou for relevant remarks and discussions about this and related points.}

All in all, the renormalized bulk+flavor brane action in the probe approximation reads
\be
S_{E}^{ren} = -\frac12 a{\cal V}\left[ 1 - \frac{48}{7}\epsilon_f \frac{\pi^{3/2}}{\Gamma\left(-\frac23\right)\Gamma\left(\frac16\right)}\right]\,.
\ee
Accordingly, the free energy density (and hence the pressure $p$) is (using (\ref{uomp}) and (\ref{ef}))
\be
f = -p =  -\frac{2 N_c^2 \lambda_4}{3^7 \pi^2}M^4_{KK}\left[1 - \frac47\frac{\lambda_4^2}{\pi^3}\frac{N_f}{N_c}\frac{\pi^{3/2}}{\Gamma\left(-\frac23\right)\Gamma\left(\frac16\right)}\right]\,.
\label{pconf}
\ee
%%%%%%%%%%%%%%%%%%%%%%%%%%%%%%%%%%%%%%%%%%
\subsection{The D8 counterterms in the backreacted case}
In the backreacted case, the $D8$ counterterms introduced above are not enough to cancel the divergences, due to extra terms in the bulk gravity fields. 
A counterterm action of the form (\ref{D8couterterm}) is still needed, but the coefficients $m_1, m_2$ given in (\ref{m12p}) have to be replaced by
\be
m_{1\,(b)}= - \frac{631}{5005}\,,\quad m_{2\,(b)} =  -\frac{2}{2145}\,.
\ee
One can equivalently argue that in the backreacted case one has to add to (\ref{ctdp}) novel counterterms with the same structure and novel coefficients.

All in all the backreacted renormalized on-shell action reads
\be
S_{E\,(b)}^{ren} = -\frac12 a{\cal V}\left[ 1 + \epsilon_f \left(4C_2+ \frac87\frac{\pi^{3/2}}{\Gamma\left(-\frac23\right)\Gamma\left(\frac16\right)}\right)\right]\,.
\ee
In order to express this in terms of field theory quantities, one has to take into account (\ref{u0Mkk}) when rewriting $a{\cal V}$. This way one finds
\be
f_{(b)} =  -\frac{2 N_c^2 \lambda_4}{3^7 \pi^2}M^4_{KK}\left[1 + \frac{\lambda_4^2}{12\pi^3}\frac{N_f}{N_c}\left(5A_1-A_2-\left(\frac67+\frac{2\pi}{\sqrt{3}}-\log\frac{256}{729}\right)\frac{\pi^{3/2}}{\Gamma\left(-\frac23\right)\Gamma\left(\frac16\right)}\right)\right]\,.
\ee
Notice that this is independent of $C_2$. In order for this expression to match with that obtained in the probe approximation we need to impose
\be\label{thermoconsist}
5A_1-A_2 = \left(-6+\frac{2\pi}{\sqrt{3}}-\log\frac{256}{729}\right)\frac{\pi^{3/2}}{\Gamma\left(-\frac23\right)\Gamma\left(\frac16\right)}\,.
\ee
Note that condition (\ref{A1A2}) is consistent with this constraint.

As in the unflavored case, the free energy of the system is independent of the temperature for any $T$ in the confined phase. 
Accordingly, the entropy density in that phase is zero\footnote{Since we work in the large $N_c$ limit, this actually means that the entropy density is of ${\cal O}(1)$, as expected in a confined phase.} as it is also evident from the absence of an event horizon in the metric. 
Thus, the pressure $p=-f$ equals minus the energy density: $\varepsilon = f = -p$. This can be also verified  by explicitly computing the energy using the standard holographic rules which equate it to the (renormalized) ADM energy of the background. The non-renormalized expression reads
\be\label{ADM}
E_{ADM} = -\frac{1}{k_0^2}\sqrt{|g_{tt}^{(E)}|}|_{r=\epsilon}\int d^8x \sqrt{g_{8}^{(E)}}K_9^{(E)}\,.
\ee
Here the metric components are written in the Einstein frame $ds_{(E)}^2 = e^{-\phi/2}ds^2$, the integral is done over the eight-dimensional slice at constant $t$ and $r=\epsilon$, and $K_9$ is the trace of the extrinsic curvature on that slice
\be
K_9^{(E)}= -\frac{1}{\sqrt{g_9^{(E)}}}\partial_r \left(\frac{\sqrt{g_9^{(E)}}}{\sqrt{g_{rr}^{(E)}}}\right)|_{r=\epsilon}\,,
\ee
written in terms of the nine-dimensional metric on the $t=const.$ slice. Evaluating the above expressions on our background we get
\be
E_{ADM} = -a{\cal V}\, T \left[ \frac12 + \frac{5}{3\epsilon}+\epsilon_f (h_1+h_2)\right] = T(S_E+S_{GH})\equiv  F_{\rm{non-ren}}\,.
\ee
Thus, the non-renormalized energy is precisely equal to the non-renormalized free energy, consistently with our expectations.

%%%%%%%%%%%%%%%%%%%%%%%%%%%%%%%%%%%%%%%%%%%%%
\section{The backreacted solution in the deconfined phase}\label{secdeconfinedcase}
\setcounter{equation}{0}
In the unflavored case \cite{witten} the deconfined phase of the theory is holographically dual to a black hole solution. 
The Euclideanized temporal circle shrinks at a radial position $u=u_T$ while the $x_4$ circle does not shrink any more. When the $D8$-branes have antipodal positions on the $x_4$ circle, the energetically favored configuration is that of a parallel stack of $D8$ and anti-$D8$ branes, such that chiral symmetry is restored \cite{Aharony:2006da}. 

\subsubsection*{Action and ansatz}
We consider smearing the two stacks of branes on the transverse circle and we turn on a $U(1)$ gauge field on the branes, which realizes a (diagonal) $U(1)_B$ symmetry in the dual field theory. The relevant action (with the flavor brane embedding put on-shell) reads
\bea\label{actionT}
2k_0^2 S & = & \int d^{10}x\sqrt{-g}\left[e^{-2\phi}\left(R+4(\partial\phi)^2\right) -\frac12 |F_4|^2\right] \nonumber\\ 
&& -\frac{2k_0^2 N_f T_8 M_{KK}}{\pi}\int d^{10}x\frac{\sqrt{-(g+2\pi\alpha' F)}}{\sqrt{g_{44}}} e^{-\phi}\,.
\eea
The string frame metric ansatz reads now
\be
ds^2 = - e^{2\tilde\lambda} dt^2 + e^{2\lambda}dx_{a}dx^{a}  + e^{2\lambda_s}dx_4^2 + l_s^2 e^{-2\varphi}d\rho^2+l_s^2 e^{2\nu}d\Omega_4^2\,,
\ee
where
\be
\varphi= 2\phi -3\lambda -\tilde\lambda-\lambda_s-4\nu\,.
\ee
We adopt an electric ansatz for the gauge field ($F=dA$)
\be
2\pi\alpha' A = {\cal A}_t (\rho) d t\,.
\ee
As usual the asymptotic value of the gauge field provides the chemical potential for the dual theory. 
Notice that since we look for a black hole solution where the Euclideanized temporal circle shrinks, a trivial solution of the form ${\cal A}_t=const.$ is not allowed any more. Instead, we have to require that the gauge field is zero at the black hole horizon.  

Implementation of the ansatz above (plus the usual ansatz in (\ref{witten0}) for $F_4$) gives the following 1d action
\bea
S &\equiv& {\cal V}\int d\rho\,L_{1d}= {\cal V}\int d\rho \left[-3{\dot\lambda}^2 - {\dot\lambda_s}^2- {\dot{\tilde\lambda}}^2 - 4{\dot\nu}^2 + {\dot\varphi}^2 + V\right]\,,\nonumber \\
V&=& 12 e^{-2\nu-2\varphi}-Q_c^2 e^{3\lambda+\lambda_s+\tilde\lambda-4\nu-\varphi}-Q_f e^{\frac32\lambda-\frac12\lambda_s+\frac{\tilde\lambda}{2}+2\nu-\frac{3}{2}\varphi}\sqrt{1-\frac{1}{l_s^2}e^{-2\tilde\lambda+2\varphi}{\dot{\cal A}}^2_t}\,,
\label{acdec}
\eea
which has to be supported by the zero-energy constraint (to be written with care, since the kinetic term for the gauge field is now contained in $V$).

\subsubsection*{Equations of motion}
The equation of motion for the gauge field can be readily solved, obtaining
\be
{\dot{\cal A}_t} = -\frac{l_s n e^{-\frac32(\lambda-\tilde\lambda)+\frac12\lambda_s-2\nu-\frac12\varphi}}{\sqrt{1+ n^2 e^{-3\lambda+\tilde\lambda+\lambda_s-4\nu+\varphi}}}\,,
\ee
where the dimensionless constant $n$ will be related to the charge density.

The remaining equations of motion (where again we reinsert the dilaton) then read
\bea
&&\ddot\lambda -\frac{Q_c^2}{2}e^{6\lambda+2\lambda_s+2\tilde\lambda-2\phi}=\frac{Q_f}{4}\frac{e^{6\lambda+\lambda_s+2\tilde\lambda-3\phi+8\nu}}{\sqrt{1+n^2 e^{-6\lambda-8\nu+2\phi}}}\,,\nonumber\\
&&\ddot\lambda_s -\frac{Q_c^2}{2}e^{6\lambda+2\lambda_s+2\tilde\lambda-2\phi}=- \frac{Q_f}{4}\frac{e^{6\lambda+\lambda_s+2\tilde\lambda-3\phi+8\nu}}{\sqrt{1+n^2 e^{-6\lambda-8\nu+2\phi}}}\,,\nonumber\\
&&
\ddot{\tilde\lambda} -\frac{Q_c^2}{2}e^{6\lambda+2\lambda_s+2\tilde\lambda-2\phi}=\frac{Q_f}{4}\frac{e^{6\lambda+\lambda_s+2\tilde\lambda-3\phi+8\nu}}{\sqrt{1+n^2 e^{-6\lambda-8\nu+2\phi}}}+\frac{Q_f}{2}\frac{n^2 e^{\lambda_s+2\tilde\lambda-\phi}}{\sqrt{1+n^2e^{-6\lambda-8\nu+2\phi}}}\,,\nonumber\\
&&
\ddot\nu +\frac{Q_c^2}{2}e^{6\lambda+2\lambda_s+2\tilde\lambda-2\phi}-3e^{6\lambda+2\lambda_s+2\tilde\lambda-4\phi+6\nu}=\frac{Q_f}{4}\frac{e^{6\lambda+\lambda_s+2\tilde\lambda-3\phi+8\nu}}{\sqrt{1+n^2 e^{-6\lambda-8\nu+2\phi}}}\,,\nonumber\\
&&
\ddot\phi -\frac{Q_c^2}{2}e^{6\lambda+2\lambda_s+2\tilde\lambda-2\phi}=\frac{5 Q_f }{4}\frac{e^{6\lambda+\lambda_s+2\tilde\lambda-3\phi+8\nu}}{\sqrt{1+n^2 e^{-6\lambda-8\nu+2\phi}}}+\frac{Q_f}{2}\frac{n^2 e^{\lambda_s+2\tilde\lambda-\phi}}{\sqrt{1+n^2e^{-6\lambda-8\nu+2\phi}}}\,.
\eea
Notice that these imply that
\be
\ddot\zeta=0\,,\quad{\rm where}\quad\zeta\equiv2\lambda-2\lambda_s+\tilde\lambda-\phi\,.
\label{zetaeq}
\ee
Since we work at first order in $N_f/N_c$, when plugging  back the gauge field into the equations of motion for the remaining functions, it is sufficient to evaluate it on the zero-th order unflavored solution. Introducing the radial coordinate 
\be
r \equiv a_T\rho\,,\qquad a_T\equiv \frac{\sqrt{2} Q_c u_T^3}{3R^3 g_s}= \frac{u_T^3}{l_s^3 g_s^2}\,,
\ee 
we can write the zero-th order unflavored solutions as
\bea
&&\lambda_0 (r) = f_0 (r) + \frac{3}{4} \log{\frac{u_T}{R}}\,,\nonumber \\
&&{\lambda_s}_0(r) =\lambda_0 (r)\,,\nonumber \\
&& \tilde\lambda_0(r) = f_0(r) -\frac{3}{2} r +  \frac{3}{4} \log{\frac{u_T}{R}}\,,\nonumber \\
&& \phi_0 (r) = f_0(r) + \frac{3}{4} \log{\frac{u_T}{R}}+\log g_s\,,\nonumber \\
&& \nu_0(r) = \frac{1}{3}f_0(r) + \frac{1}{4} \log{\frac{u_T}{R}}+\log\frac{R}{l_s}\,,\nonumber\\
&& {\cal A}'_t (r) = - u_T\frac{q e^{-3r}}{\sqrt{1-e^{-3r}}}\frac{1}{\sqrt{1+q^2 (1-e^{-3r})^{5/3}}}\,,
\eea
with
\be
f_0(r) = -\frac{1}{4}\log\left[1-e^{-3r}\right]\,,
\ee
and
\be
q = \frac{g_s l_s^4}{R^{3/2}u_T^{5/2}}n\,.
\ee
Notice that, consistently with the all-order equation (\ref{zetaeq})
\be
\zeta_0\equiv2\lambda_0-2{\lambda_s}_0+\tilde\lambda_0-\phi_0=-\log g_s -\frac32 r\,.
\label{zeta0}
\ee
As we have done in section \ref{secconfinedcase}, we solve the equations above in a perturbative expansion, defining for each field
\be
\Psi=\Psi_0 +  \epsilon_{f\,T }\Psi_1 + {\cal O}(\epsilon_T^2)\,,
\ee
with
\be
\epsilon_{f\,T} \equiv \frac{R^{3/2} u_T^{1/2} g_s}{l_s^2}Q_f = \epsilon_f \sqrt{\frac{u_T}{u_0}}=\frac{\lambda_4^2}{12\pi^3}\frac{2\pi T}{M_{KK}}\frac{N_f}{N_c}\ll 1\,,
\label{epst}
\ee
where we have used the zero-th order relations between $u_0\,,u_T$ and $M_{KK}\,,T$.\footnote{Since $\lambda_4, \epsilon_f, {\cal A}_t$ are already ``first order quantities'' in $N_f/N_c$, in their expressions one can use the leading relations between $u_0, u_T$ and $M_{KK}, T$.} 
The above definition is such that $\epsilon_{f\,T}=\epsilon_f$ at the phase transition. 
Moreover, it suggests a definition of a ``running" flavor coupling $\epsilon_f(u)\sim\sqrt{u}\sim E$ where $E$ is the field theory RG energy scale.\footnote{The beta function for this running coupling is thus $\beta(\epsilon_f(E))=\epsilon_f(E)$.}

The equations of motion we need to solve are thus (derivatives are w.r.t. $r$)
\bea\label{eqsq}
&& \lambda_1''-\frac92\frac{e^{-3r}}{(1-e^{-3r})^2}(3\lambda_1+\lambda_{s\,1}+\tilde\lambda_1-\phi_1)=\frac14\frac{e^{-3r}}{(1-e^{-3r})^{13/6}}\frac{1}{\sqrt{1+q^2 (1-e^{-3r})^{5/3}}}\,, \nonumber\\
&& \lambda_{s\,1}''-\frac92\frac{e^{-3r}}{(1-e^{-3r})^2}(3\lambda_1+\lambda_{s\,1}+\tilde\lambda_1-\phi_1)=-\frac14\frac{e^{-3r}}{(1-e^{-3r})^{13/6}}\frac{1}{\sqrt{1+q^2 (1-e^{-3r})^{5/3}}}\,,\nonumber\\
&& \tilde\lambda_{1}''-\frac92\frac{e^{-3r}}{(1-e^{-3r})^2}(3\lambda_1+\lambda_{s\,1}+\tilde\lambda_1-\phi_1)=\frac14\frac{e^{-3r}}{(1-e^{-3r})^{13/6}}\frac{1}{\sqrt{1+q^2 (1-e^{-3r})^{5/3}}}+\nonumber\\
&& \qquad \qquad \qquad \qquad \qquad \qquad\qquad \qquad +\frac12\frac{e^{-3r}}{\sqrt{1-e^{-3r}}}\frac{q^2}{\sqrt{1+q^2(1-e^{-3r})^{5/3}}}\,,\nonumber\\
&& \phi_1''-\frac92\frac{e^{-3r}}{(1-e^{-3r})^2}(3\lambda_1+\lambda_{s\,1}+\tilde\lambda_1-\phi_1)=\frac54\frac{e^{-3r}}{(1-e^{-3r})^{13/6}}\frac{1}{\sqrt{1+q^2 (1-e^{-3r})^{5/3}}}+\nonumber \\
&& \qquad \qquad \qquad \qquad \qquad \qquad\qquad \qquad  +\frac12\frac{e^{-3r}}{\sqrt{1-e^{-3r}}}\frac{q^2}{\sqrt{1+q^2(1-e^{-3r})^{5/3}}}\nonumber \,,\\
&& \nu_1''-\frac32\frac{e^{-3r}}{(1-e^{-3r})^2}(3\lambda_1+\lambda_{s\,1}+\tilde\lambda_1-5\phi_1+12\nu_1)= \nonumber\\
&& \qquad\qquad\qquad\qquad\qquad\qquad \qquad\qquad +\frac14\frac{e^{-3r}}{(1-e^{-3r})^{13/6}}\frac{1}{\sqrt{1+q^2 (1-e^{-3r})^{5/3}}}\, .
\eea
In this paper we do not attempt solving these equations exactly in $q$. Instead we focus on the small charge case, keeping only the leading $q^2$ terms in an expansion.

\subsection{The small charge solutions}
The simplest expression is that of the gauge field which, in this limit, reads
\be
{\cal A}_t = \frac23 q\, u_T \left(1-\sqrt{1-e^{-3r}}\right)\,.
\label{asq}
\ee
The remaining solutions read
\bea
&&\lambda_1 = \frac{f}{28} - \frac{3}{16} q^2 g +y -\frac14 (a_2-a_1-a_3)-\frac14 (b_2-b_1-b_3)r \,,\nonumber\\
&& \lambda_{s\,1}=\lambda_1-\frac{f}{21} + \frac{q^2}{4}g- a_1 - b_1 r\,,\nonumber\\
&& \tilde\lambda_1 = \lambda_1 + \frac{q^2}{2}g- a_3-b_3 r\,,\nonumber\\
&& \phi_1 = \lambda_1 + \frac{2}{21}f - a_2 -b_2 r\,,\nonumber \\
&& \nu_1 = \frac{11}{252}f - \frac{q^2}{16} g + w\,,
\eea
where
\bea
&& f =\frac{6}{\sqrt[6]{1-e^{-3 r}}}+\sqrt{3} \tan
   ^{-1}\left(\frac{2 \sqrt[6]{1-e^{-3 r}}-1}{\sqrt{3}}\right)+\sqrt{3} \tan
   ^{-1}\left(\frac{2 \sqrt[6]{1-e^{-3 r}}+1}{\sqrt{3}}\right)+\nonumber \\
   && \qquad -2 \tanh
   ^{-1}\left(\sqrt[6]{1-e^{-3 r}}\right)-\coth ^{-1}\left(\sqrt[6]{1-e^{-3
   r}}+\frac{1}{\sqrt[6]{1-e^{-3 r}}}\right)\,,\nonumber\\
&& g = \frac{4}{9} \log \left(e^{-3 r/2} \sqrt{e^{3 r}-1}+1\right)-\frac{4}{9} e^{-3
   r/2} \sqrt{e^{3 r}-1}\,,\nonumber\\
&& y = c_2 - \left(c_1+(1+\frac32 r) c_2\right)\coth\left(\frac{3r}{2}\right) + q^2 j+ z\,, \nonumber \\   
&& w = 2 m_2 - (m_1 + (2+3r) m_2)\coth\left(\frac{3r}{2}\right)+\frac{1}{12}(a_1-5a_2+a_3+b_1 r -5 b_2 r + b_3 r) +\nonumber \\
&& \qquad +\frac53 z - q^2 j  \nonumber \,,\\
&& j = \frac{1}{72} \left(4 \sqrt{1-e^{-3 r}}+\left(-9 r+6 \sqrt{1-e^{-3 r}}-6 \log
   \left(\sqrt{1-e^{-3 r}}+1\right)\right) \coth \left(\frac{3
   r}{2}\right)\right)\,,\nonumber\\
&& z= \frac{3 e^{3 r} \left(1-e^{-3 r}\right)^{5/6}-\frac{\sqrt{3}}{2} \left(e^{3 r}+1\right)
   \left[\tan ^{-1}\left(\frac{2 \sqrt[6]{1-e^{-3
   r}}-1}{\sqrt{3}}\right)+\tan ^{-1}\left(\frac{2 \sqrt[6]{1-e^{-3
   r}}+1}{\sqrt{3}}\right)\right]}{546 \left(e^{3 r}-1\right)}+\nonumber\\
&& \qquad +\frac{1}{2} \left(e^{3 r}+1\right)\frac{2 \tanh ^{-1}\left(\sqrt[6]{1-e^{-3
   r}}\right)+\coth ^{-1}\left(\sqrt[6]{1-e^{-3 r}}+\frac{1}{\sqrt[6]{1-e^{-3
   r}}}\right)}{546 \left(e^{3 r}-1\right)} \,.
\eea
In the small-$q$ case, the zero-energy condition reads
\be
-3{\dot\lambda}^2 - {\dot\lambda_s}^2- {\dot{\tilde\lambda}}^2 - 4{\dot\nu}^2 + {\dot\varphi}^2 +\frac{Q_f}{2 l_s^2}e^{\frac32\lambda-\frac{1}{2}\lambda_s{\bf -\frac{3}{2}\tilde\lambda}+2\nu{\bf +\frac{1}{2}\varphi}}{\dot{\cal A}}^2_t - {\cal P} =0\,,
\ee
with
\be
{\cal P}= 12 e^{-2\nu-2\varphi}-Q_c^2 e^{3\lambda+\lambda_s+\tilde\lambda-4\nu-\varphi}-Q_f e^{\frac32\lambda-\frac12\lambda_s+\frac{\tilde\lambda}{2}+2\nu-\frac{3}{2}\varphi}\,.
\ee
This condition can be satisfied if
\be
-2b_1 - 2b_2+10 b_3 + 3(-12 c_2 -48 m_2 + q^2) = 0\,.
\label{zet}
\ee
The near-horizon behavior of the functions reads (here $x=e^{-3r/2}\rightarrow 0$ near the horizon)
\bea
&&\lambda_1 = \frac{819 a_1-819 a_2+819 a_3+3 \log (x) \left(-182
   b_1+182 b_2-182 b_3+1092 c_2+91 q^2+76\right)}{3276}\nonumber \\
&&\qquad + \frac{-3276
   c_1-182 q^2 (\log (8)-4)+57 \sqrt{3} \pi +720-114 \log (12)-57 \log
   (3)}{3276} + {\cal O}(x^2)\,,\nonumber \\
&&\tilde\lambda_1 = \frac{819 a_1-819 a_2-2457 a_3+3 \log (x) \left(-182
   b_1+182 b_2+546 b_3+1092 c_2+91 q^2+76\right)}{3276}\nonumber \\
&&\qquad  +\frac{-3276
   c_1+182 q^2 \log (2)+57 \sqrt{3} \pi +720-114 \log (12)-57 \log
   (3)}{3276} + {\cal O}(x^2)\,,\nonumber \\
&& \lambda_{s\,1} = \frac{-2457 a_1-819 a_2+819 a_3+21 \log (x) \left(78
   b_1+26 b_2-26 b_3+156 c_2+13 q^2-4\right)}{3276}\nonumber \\
&&\qquad  +\frac{  -3276
   c_1-182 q^2 (\log (2)-2)-21 \sqrt{3} \pi +3 (7 \log (432)-72)}{3276} + {\cal O}(x^2)\,,\\
&& \phi_1 = \frac{819 a_1-4095 a_2+819 a_3 + 273 \log (x) \left((852/273) + q^2- 2( b_1 -5 b_2 + b_3-6 c_2\right)}{3276}\nonumber \\
&&\qquad  +\frac{2592-3276 c_1-182 q^2
   (\log (8)-4) -852 \log \left(2\right)+213 \sqrt{3} \pi -639
   \log (3)}{3276}+{\cal O}(x^2)\,,\nonumber \\
&& \nu_1 = \frac{273 a_1-1365 a_2+273 a_3+\log (x) \left(276-182
   b_1+910 b_2-182 b_3+6552 m_2-273 q^2\right)}{3276}\nonumber \\
&&\qquad +\frac{-3276
   m_1+182 q^2 (\log (2)-2)+69 \sqrt{3} \pi +888-138 \log (12)-69 \log
   (3)}{3276} + {\cal O}(x^2)\,.\nonumber
\eea            
Regularity at the tip of the $(t,r)$ Euclideanized cigar leads to the constraints
\be
b_1 = \frac17\,,\quad b_2 =-\frac27\,,\quad c_2 =\frac{1}{546}+\frac{b_3}{6}-\frac{q^2}{12}\,,\quad m_2 = \frac{5}{3276}+\frac{b_3}{36}+\frac{q^2}{24}\,,
\label{irT}
\ee
which automatically fulfill the zero-energy constraint (\ref{zet}). These conditions can be further reinforced by requiring all the logarithmically divergent terms to disappear by imposing\footnote{As in the confined case, this is not strictly necessary.}
\be
b_3=0\,.
\label{b3}
\ee
Notice moreover that, once the regularity conditions (\ref{irT}) are satisfied, it follows that
\be
\zeta_1 = 2a_1+a_2-a_3 + b_3 r\,,
\label{zeta1}
\ee
consistently with the general equation (\ref{zetaeq}). Comparing (\ref{zeta1}) with (\ref{zeta0}) we see that the combination $2a_1 +a_2 -a_3$ can be read as a correction to the string coupling $g_s$, while the constant $b_3$ can be seen as a flavor-dependent rescaling of the radial variable $r$. As in the confined case, thus, a possible all-order choice on the integration constants would be to set $\zeta=\zeta_0$ so that, in the case at hand $b_3 =0$, $a_3= 2a_1 + a_2$.

The UV behavior ($x\rightarrow1$) of the functions is given by
\bea
&&\lambda_1= -\frac{c_1+c_2}{1-x}+\frac{101}{455(2)^{1/6}(1-x)^{1/6}} + \frac14(a_1-a_2+a_3+2(c_1+c_2)) + {\cal O}(z^{1/6})\,, \\
&& \lambda_{s\,1} = -\frac{c_1+c_2}{1-x}-\frac{29}{455(2)^{1/6}(1-x)^{1/6}} + \frac14(-3a_1-a_2+a_3+2(c_1+c_2)) + {\cal O}(z^{1/6})\,,\nonumber \\
&& \tilde\lambda_1 = -\frac{c_1+c_2}{1-x}+\frac{101}{455(2)^{1/6}(1-x)^{1/6}} + \frac14(a_1-a_2-3a_3+2(c_1+c_2)) + {\cal O}(z^{1/6})\,,\nonumber \\
&& \phi_1 = -\frac{c_1+c_2}{1-x}+\frac{361}{455(2)^{1/6}(1-x)^{1/6}} + \frac14(a_1-5a_2+a_3+2(c_1+c_2)) + {\cal O}(z^{1/6})\,,\nonumber \\
&& \nu_1 = -\frac{m_1+2m_2}{1-x}+\frac{25}{91 (2)^{1/6}(1-x)^{1/6}} + \frac{1}{12}(a_1-5a_2+a_3+6m_1)+m_2 + {\cal O}(z^{1/6})\,.\nonumber
\eea
Requiring the leading divergences to be absent amounts on having
\be
c_1= - c_2\,,\qquad m_1 = - 2m_2\,.
\label{uvtsources}
\ee
The constant terms can be eliminated by further imposing 
\be
a_1=a_2=a_3=0\,,
\label{uvtvevs}
\ee
which would be consistent with the all-order choice of the integration constants $\zeta=\zeta_0$.
As before we work in the more general case where the conditions (\ref{irT}), (\ref{uvtsources}) hold. 
We will sometimes specialize to the more restricted cases where (\ref{b3}), (\ref{uvtvevs}) are also satisfied.

%%%%%%%%%%%%%%%%%%%%%%%%%%%%%
\section{Thermodynamics}\label{secthermo}
In this section we present the thermodynamic of the theory computed from the gravitational background of section \ref{secdeconfinedcase}.\footnote{See \cite{ihl} for similar studies in a non-supersymmetric model based on $D3-D7-\bar{D}7$-branes.}
Let us begin by first relating our dimensionless parameter $q$ with the quark chemical potential and density. 
Using the holographic relation between the asymptotic value of the original field $A_t$ and the chemical potential $\mu$ we get, from (\ref{asq}), to leading order in the $N_f/N_c$ expansion
\be
\mu = \frac{1}{3\pi}\frac{q\, u_T}{l_s^2}= \frac{8\pi}{27}q\lambda_4\frac{T^2}{M_{KK}}= \frac{4}{27}q\lambda_{4\,T}T \,,
\label{muq}
\ee
where we have defined the running 't Hooft coupling
\be
\lambda_{4\,T} \equiv \lambda_4 \frac{2\pi T}{M_{KK}}\,.
\label{runtof}
\ee
The quark charge density is obtained from the holographic relation
\be
n_q = -g_s^{-1}\frac{\delta {\cal L}_{(5)}}{\delta F_{t \hat\rho}}\,,
\ee
where ${\cal L}_{(5)}$ is the five dimensional Lagrangian density (obtained from the original 10d smeared one after reduction over $x_4$ and $S^4$) and $\hat\rho= l_s \rho$. An explicit computation gives
\be\label{nq}
n_q = \frac{32\pi}{729}q N_c N_f \lambda_4^2 \frac{T^5}{M_{KK}^2}= \frac{8}{729\pi}q N_c N_f \lambda_{4\,T}^2 T^3\,.
\ee
Notice that, as usual, the baryon charge density is defined as
\be
n_b = \frac{n_q}{N_c}\,.
\ee
From expressions (\ref{muq},\ref{nq}) it follows that when performing calculations in the grand-canonical ensemble, where the chemical potential is held fixed, we should consider that $q\sim T^{-2}$. Instead, in the canonical ensemble $q\sim T^{-5}$.

Requiring the holographic computation of the thermodynamic observables (using the renormalized on-shell action for the pressure and the renormalized ADM energy for the energy density) to give consistent thermodynamic relations imposes the constraint 
\be
5 a_3 = a_1 + a_2\,.
\label{relas}
\ee
In the following presentation this additional constraint is implemented. 

For the Euclideanized black hole metric to be regular at the horizon, the relation
\be
\frac{u_T}{R^3} = \frac49 (2\pi T)^2 \left[ 1 +\frac29 \epsilon_{f\,T}\left(1 - b_3+\frac{q^2}{2}\right)\right]
\label{utT1}
\ee
must hold.
Using the Bekenstein-Hawking formula one gets the entropy density
\be
s = \frac{256 N_c^2 \pi^4 \lambda_4}{729 M_{KK}^2}T^5\left[1+\frac{2}{3}\epsilon_{f\,T}\left(1+\frac{q^2}{2}\right)\right]\,,
\ee
from which we can deduce the Helmholtz free energy density using $s=-(\partial f/\partial T)$ (canonical ensemble), taking into account the $T$-dependence of $\epsilon_{f\,T}$ as in (\ref{epst}) and the $T$-dependence of $q$ in the canonical ensemble 
\be\label{freeen}
f = - \frac16 \left(\frac{256 N_c^2 \pi^4 \lambda_4}{729 M_{KK}^2}\right) T^6 \left[1+\frac47 \epsilon_{f\,T}\left(1-\frac76 q^2\right)\right]\,,
\ee
and, using $\varepsilon = Ts +f$, the energy density
\be\label{en}
\varepsilon= \frac56 \left(\frac{256 N_c^2 \pi^4 \lambda_4}{729 M_{KK}^2}\right) T^6 \left[1+\frac{24}{35} \epsilon_{f\,T}\left(1+\frac79 q^2\right)\right]\,.
\ee
The value of the energy density (\ref{en}) also follows from the explicit computation of the renormalized ADM energy (\ref{ADM}) of the background.
The heat capacity at fixed quark density reads
\be
c_{V,n}=\left(\frac{\partial\varepsilon}{\partial T}\right)_{V,n} = 5\left(\frac{256 N_c^2 \pi^4 \lambda_4}{729 M_{KK}^2}\right) T^5\left[1+\frac45\epsilon_{f\,T}\left(1-\frac13 q^2\right)\right]\,.
\ee
Note that in the formulas above $M_{KK}$ comes directly from the length of the $S_{x_4}$ circle, so it represents the flavored energy scale.

Passing to the grand-canonical ensemble, we first notice that
\be
\mu n_q = \frac29 \left(\frac{256 N_c^2 \pi^4 \lambda_4}{729 M_{KK}^2}\right)\epsilon_{fT}\, q^2 T^6\,,
\ee
so that the Gibbs free energy density reads
\be
\omega = - p = -\frac16  \left(\frac{256 N_c^2 \pi^4 \lambda_4}{729 M_{KK}^2}\right) T^6 \left[1+\frac47\epsilon_{f\,T}\left(1+\frac76 q^2\right)\right]\,,
\label{gibbs}
\ee
where $p$ is the pressure. The above expression satisfies the thermodynamic relations $\omega = f - \mu n_q$ and $s=-\partial\omega/\partial T$ (at fixed $\mu$). Moreover it satisfies
\be
\frac{\delta\omega}{\delta\mu}= - n_b\,.
\ee
The trace of the stress energy tensor reads
\be
\varepsilon-3p =\frac13\left(\frac{256 N_c^2 \pi^4 \lambda_4}{729 M_{KK}^2}\right) T^6 \left[1+\frac67\epsilon_{f\,T}\left(1+\frac{7}{18} q^2\right)\right]\,.
\ee
The heat capacity at fixed chemical potential is
\be
c_{V,\mu}=\left(\frac{\partial\varepsilon}{\partial T}\right)_{V,\mu}=5\left(\frac{256 N_c^2 \pi^4 \lambda_4}{729 M_{KK}^2}\right) T^5 \left[1+\frac45\epsilon_{f\,T}\left(1+\frac{1}{3} q^2\right)\right]\,,
\ee
so that the squared speed of sound reads
\be
c_s^2 = \frac{s}{c_{V,\mu}} = \frac15\left[1-\frac{2}{15}\epsilon_{f\,T}\left(1-\frac{1}{2} q^2\right)\right]\,.
\ee
Note that the dependence on the integration constants dropped out of the thermodynamical observables, as expected.
The formulas above agree with those which can be obtained in the probe approximation, see e.g. \cite{Kim:2006gp,parnakim}, though a covariant treatment of the holographic renormalization procedure needed in that case was missed in those works.

The holographic computation of the free energy requires a careful treatment of the boundary (counter)-terms. Actually these terms are the same as in section \ref{holren}, with the only difference being on the coefficients.
%%%%%%%%%%%%%%%%%%%%%%%%%%%%%%%%%%%%
\subsection{Holographic renormalization}
\label{holrendec}
As for the confined solution,  the on-shell action (\ref{acdec}) (and so the Gibbs free energy\footnote{The Helmholtz free energy $F$ is holographically related to the Legendre transformed on-shell action, which, once reduced to a radial integral is defined as in (\ref{acdec}) with $L_{1d}$ replaced by its Legendre transform ${\tilde L}_{1d} = L_{1d} - (\delta L_{1d}/\delta A'_{t}) A'_t$, where the last term has to be evaluated on the solution of the equation of motion for the gauge field.} $\Omega$) in the deconfined case is a divergent quantity and must be renormalized by appropriate counterterms
\be
S_{E}^{ren} = (S_{E} + {S}_{GH})+S_{c.t.}^{bulk}+S_{c.t.}^{D8}\,,
\ee
where the functional form of each term is as in formulas (\ref{actionT}), (\ref{gibbonshawking}), (\ref{bulkcounterterm}), (\ref{D8couterterm}) respectively. 
On the deconfined solution they read
\bea
&& S_{E}= - a_T{\cal V}\left[\frac94 -\frac{3}{2\epsilon}+\epsilon_{f\,T} \left(\frac32 b_3 + \frac{25}{14} + \frac34 q^2 -\frac{823}{1365 (3)^{1/6}\epsilon^{7/6}}-\frac{955}{924(3)^{1/6}\epsilon^{1/6}} \right)\right]\,,\nonumber\\
&& S_{GH} =  - a_T{\cal V}\left[-\frac74 +\frac{19}{6\epsilon}+\epsilon_{f\,T} \left(-\frac76 b_3 - \frac{11}{6} - \frac{7}{12} q^2+\frac{25}{39 (3)^{1/6}\epsilon^{7/6}}+\frac{137}{132(3)^{1/6}\epsilon^{1/6}} \right)\right]\,,\nonumber\\
&& S^{bulk}_{c.t.} =  a_T{\cal V}\left[\frac{5}{3\epsilon}+\epsilon_{f\,T} \left( \frac{14}{117 (3)^{1/6}\epsilon^{7/6}}+\frac{5}{198(3)^{1/6}\epsilon^{1/6}}\right)\right]\,,
\eea
where now 
\be
a_T{\cal V}=\frac{1}{2k_0^2g_s^2}\frac{V_3}{T}\frac{2\pi}{M_{KK}}V_{S^4}u_T^3\,.
\ee

As in the confined case, we see that the ``bulk counterterm'' $S^{bulk}_{c.t.}$ only cancels the ${\cal O}(\epsilon_{f\,T}^0)$ divergences as in the unflavored case. 
In order to cancel the remaining divergences, we need to introduce additional counterterms related to the $D8$-branes. 
Another possibility, often considered in the literature, would be to cancel the divergences by subtracting to the on-shell value of $S_E+S_{GH}$, the value of the same combination on some reference background. 
In our case a natural choice would be keeping as reference background the one corresponding to the confined phase. 
However, by computing the difference between $S_E+S_{GH}$ here and the related on-shell value obtained in section \ref{holren} one discovers that the ${\cal O}(\epsilon_f)$ divergences do not cancel.\footnote{As usual, care has to be taken when comparing on-shell values of the action on different backgrounds: in particular one has eventually to rescale the coordinates in order for the backgrounds to coincide on the asymptotic slice at $r=\epsilon$.} 
This is true also in the probe approximation: the divergences related to the on-shell $D8$-brane actions do not cancel among deconfined and confined solutions. 
This is the main reason why, in this work, we perform renormalization by adding counterterms separately in the two phases.

In the probe approximation the on-shell value of the DBI action for the $D8$-branes in the small-$q$ limit reads
\be
S_{D8}  = a_T\,{\cal V}\, \epsilon_{f\,T}\left[\frac{2}{21(3)^{1/6}\epsilon^{7/6}}+\frac{1}{6(3)^{1/6}\epsilon^{1/6}}-\frac27-\frac13 q^2\right]\,.
\label{sd8dec}
\ee
It is relevant to notice that there are no charge-dependent divergences. 
This is actually true for any $q$ (the corresponding on-shell action for the probe branes can be easily computed in this case) and it was already noticed in the past (see e.g. \cite{parnakim}). 
Thus the covariant counterterms we have introduced are enough to cancel the divergences in the charged case too. 
Notice instead that the divergences do not cancel when (carefully) subtracting the on-shell $D8$-actions (\ref{sd8dec}), (\ref{sd8con}) in the deconfined and confined phases.

All in all, when flavors are added in the probe approximation, the additional counterterm is the one in (\ref{D8couterterm}) with 
\be
m_1=\frac17\,, \qquad  m_2= \frac{1}{14}.
\ee
 The counterterm for the $D8$-branes in standard eight dimensional form reads\footnote{Again, remember that in our notation $T_8=(2\pi)^{-8}\alpha'^{-9/2}$.}
\be 
S_{c.t.}^{D8} = 2 N_f T_8 \int d^8x \sqrt{h_8}\left[ \frac{2}{7}\frac{R}{g_s^{1/3}}\, e^{-2\phi/3}-\frac{2}{7}\frac{R^2}{g_s^{2/3}}\, e^{-\phi/3}\left( K_9  -\frac83 n\cdot\nabla\phi \right)\right]\,.
\ee 

In the backreacted case, the divergences are canceled provided
\be
m_{1\,(b)}= -\frac{607}{5005},\qquad  m_{2\,(b)}=\frac{ 4}{15015}\,.
\ee
Using (\ref{utT1}), we get that, once the relation (\ref{relas}) is implemented, the Gibbs free energy density reads
\be
\omega= \frac{S_E^{ren} T}{V_3}=-\frac16  \left(\frac{256 N_c^2 \pi^4 \lambda_4}{729 M_{KK}^2}\right) T^6 \left[1+\frac47\epsilon_{f\,T}\left(1+\frac76 q^2\right)\right]\,,
\ee
which precisely matches with (\ref{gibbs}).

The divergences appearing in the computation of the ADM energy are exactly the ones above.
Thus, the counterterms needed to renormalize it are the same. 

%%%%%%%%%%%%%%%%%%%%%%%%%%%%%%%%%%%%%%%
\section{Phase diagrams}\label{secphases}

In this section we discuss the phase diagrams of the holographic model, comparing them to the lattice results for QCD. 
We will first focus on the finite temperature, finite baryon chemical potential setup, and then we will also consider the model at imaginary chemical potential and at finite $\theta$ angle.

%%%%%%%%%%%%%%%%%%%%%%%%%%%%%%%%%%%%%%%%%%%
\subsection{The critical temperature}\label{sectemperature}
We want to quantify the effects of the flavor fields on the critical temperature $T_c$ at which the first order transition between the deconfined and the confined phase happens.
To get $T_c$ we just need to solve the equation $p_{conf}=p_{deconf}(T_c)$ which, using (\ref{pconf}) and (\ref{gibbs}) gives, to first order in $N_f/N_c$
\be\label{tici}
\frac{2\pi T_c}{M_{KK}} = 1-\frac{1}{126 \pi^3}\lambda_4^2\frac{N_f}{N_c}\left(1+\frac{12\pi^{3/2}}{\Gamma\left(-\frac23\right)\Gamma\left(\frac16\right)}\right)-\frac{27}{16\pi}\frac{N_f}{N_c}\frac{\mu^2}{M_{KK}^2}\,,
\ee
where we have chosen to work at fixed quark chemical potential $\mu$, using formula (\ref{muq}).
It is interesting to notice that, for any $N_f$, the contribution of the baryon chemical potential is quadratic and it is such that $T_c$ decreases when $\mu$ increases. 
This is in agreement with lattice QCD results.

Instead, since
\be
\left(1+\frac{12\pi^{3/2}}{\Gamma\left(-\frac23\right)\Gamma\left(\frac16\right)}\right)\approx -1.987\,,
\ee
one might naively conclude that, at zero chemical potential, the effect of the flavors is that of increasing the critical temperature (see also \cite{ballon}). 
Of course this is a scheme-dependent statement which means that, as in QCD \cite{mpl}, it depends on which appropriately chosen physical observable (e.g. the string tension or the $\rho$-meson mass) or suitable UV mass scale (as in \cite{qcddari} or \cite{shuryak}) is held fixed when comparing theories with different numbers of flavors.
In fact, while the chemical potential is a parameter of a given theory and so one can unambiguously compare the theory at different values of $\mu$, this is not the case for $N_f$.
By varying the number of flavors one is changing the theory: comparison of two theories requires a scheme and there is no a-priori ``correct scheme'' - it depends on the different physical effects one wants to investigate.
See \cite{Bigazzi:2009gu,Bigazzi:2009bk,Bigazzi:2011it} for some examples and discussions in the holographic context.

In the present case the behavior of $T_c$ with $N_f$ changes qualitatively by changing comparison scheme.
For example, one of the comparison schemes used in lattice literature consists in fixing the coupling at a certain energy scale.
In our case, this produces a shift in $u_{0}$, giving an additional term in (\ref{tici}) which depends on the solution's parameters - the sources of gauge invariant operators. 
Now, for some choices of these parameters, the overall effect of the flavors is still that of increasing the critical temperature.
But for large enough values of parameters the effect is the opposite, provided the energy at which we fix the coupling is appropriately chosen.

Analogously, while the comparison scheme which fixes the mass of the $\rho$ meson gives an increasing $T_c$ with $N_f$, one can find a simple scheme where  $T_c$ decreases with $N_f$, by fixing the ratio of the string tension with $M_{KK}^2$.
Thus, not having a particular reason to prefer a comparison scheme w.r.t. another one in the present discussion, we will not consider the dependence of $T_c$ on $N_f$ furthermore in the following discussion of the phase diagrams.

%%%%%%%%%%%%%%%%%%%%%%%%%%%%%%%%%%%%%%%%%%
\subsection{Finite $\theta$ and imaginary chemical potential}

An interesting parallel between the phase diagram of (quenched) QCD at finite $\theta_B\equiv \mu_B/T_c(0)$ (where $\mu_{B}$ is the imaginary baryon chemical potential\footnote{Notice that here $\mu_B= N_c\mu_q$ where $\mu_q$ is the imaginary chemical potential for quark number.} and $T_c(0)$ the critical temperature at $\mu_B=0$) and that of the pure Yang-Mills theory at finite $\theta$ angle has been traced in \cite{massimo2}. 
Let us describe this relation and its realization in the holographic model at hand.

When $\theta$ is turned on, the lattice QCD partition function displays imaginary terms which prevent Monte Carlo simulations to numerically converge (see e.g. \cite{vicari} for a review). This ``sign problem'' (analogous to the one encountered at finite, real, baryon chemical potential) can be avoided if the theory is formulated at imaginary values of  $\theta$.
In recent years lattice studies of  the QCD phase diagram at finite imaginary chemical potential \cite{imchem} or $\theta$ parameter \cite{imtheta} have been performed. 
Assuming the (free energy of the) theory to be analytic around $\theta_B=0$ or $\theta=0$ one can perform an analytic continuation towards real values (and a further continuation to the continuum limit) to get information on the phase diagram of the theory at small real values of these parameters.

Using this trick, for example, the effect of a finite small $\theta$ parameter on the deconfining temperature $T_c$ of pure $SU(3)$ Yang-Mills was considered in \cite{massimo1}. The result was found to be in qualitative agreement with the following formula
\be
\frac{T_c(\theta)}{T_c(0)} = 1 - r_{\theta}\frac{\theta^2}{N_c^2}+ {\cal O}(\theta^4/N_c^2)\,,
\label{quad}
\ee
arising from a large $N_c$ estimate in the $SU(N_c)$ Yang-Mills case. 
For $N_c=3$ lattice results give $T_c(\theta)/T_c(0) = 1- R_{\theta}\,\theta^2 + {\cal O}(\theta^4)$, with $R_{\theta}\sim 0.0175$, while large $N_c$ arguments suggest $r_{\theta}\sim 0.25$ \cite{massimo1}. In both cases, the deconfining temperature is found to be a quadratically decreasing function of $\theta$. 

Going beyond the small $\theta$ regime is not possible on the lattice at the moment, due to the above mentioned sign problem. 
However, at least in the large-$N_c$ limit of Yang-Mills, it is possible to figure out what is the structure of $\theta$-dependent observables just by using simple scaling arguments. 
In 't Hooft's large $N_c$ limit, $\theta$-dependence can remain non trivial provided $\theta/N_c$ is held fixed.  
In order to reconcile this condition with, say, the periodicity condition $\varepsilon(\theta)=\varepsilon(\theta+2\pi k)$ on the vacuum energy density of the theory, Witten proposed that the latter (which in turn should have an absolute minimum at $\theta=0$) should have the form \cite{wittold, Witten:1998uka}
\be
\varepsilon(\theta) = b N_c^2\, {\rm min_k} \left(\frac{\theta+2\pi k}{N_c}\right)^2 + {\cal O}(1/N_c)\,,
\label{wit}
\ee
where $b$ is a model-dependent dimensional coefficient (independent on $N_c$) and $k$ is an integer. The vacuum energy is thus expected to have discontinuities at $\theta=(2k+1)\pi$. At each of these points (where the model is barely CP-invariant), a first order quantum phase transition occurs between two different phases, spontaneously breaking CP symmetry.

While the above structure cannot be verified yet on the lattice, formula (\ref{wit}) found an explicit realization in \cite{Witten:1998uka}. There, the computation was done, using the holographic gauge/gravity duality, for the unflavored version \cite{witten} of the model we have focused on in this work. 

\begin{figure}[t]
\centering
\includegraphics[width=80mm]{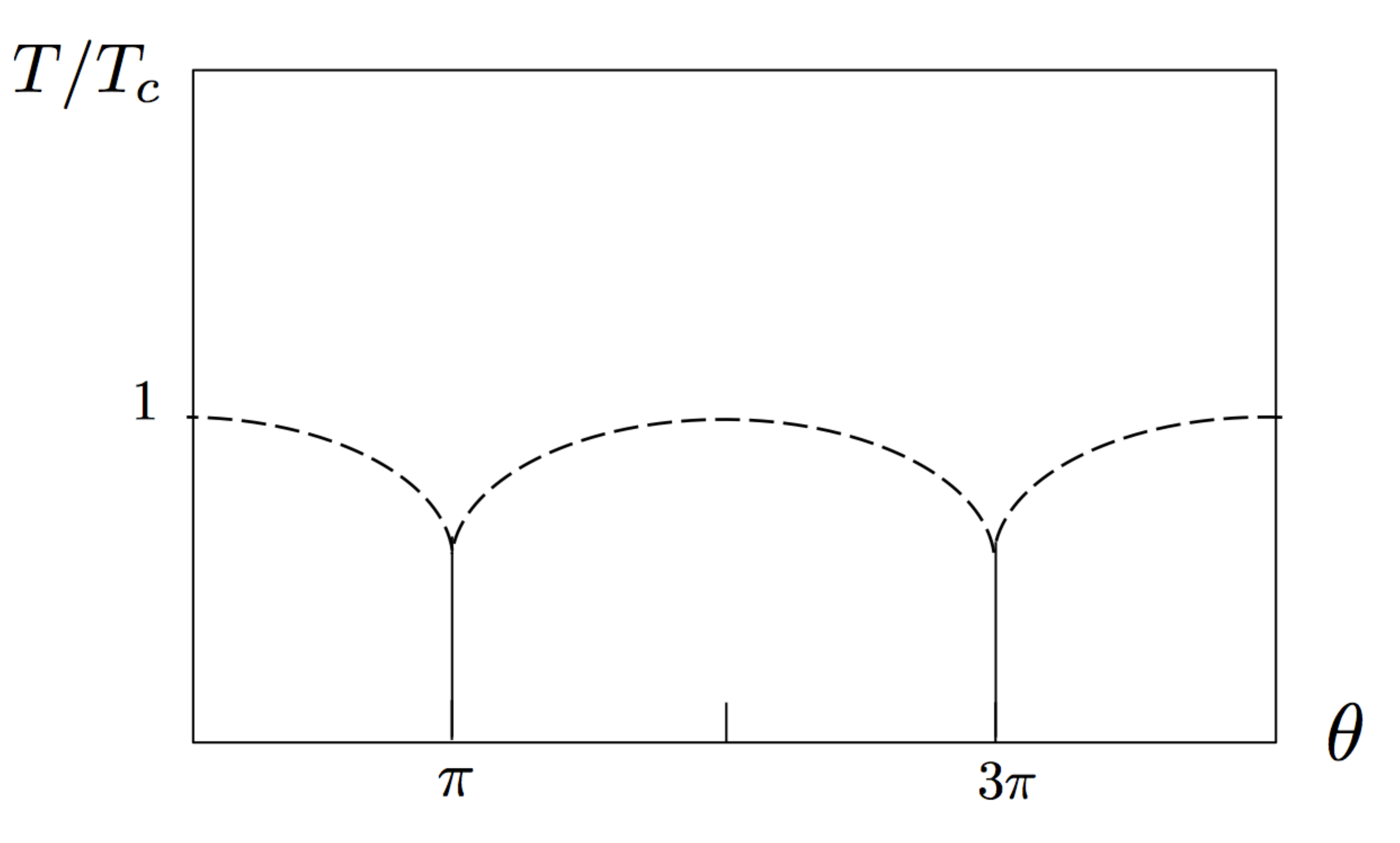}
\includegraphics[width=80mm]{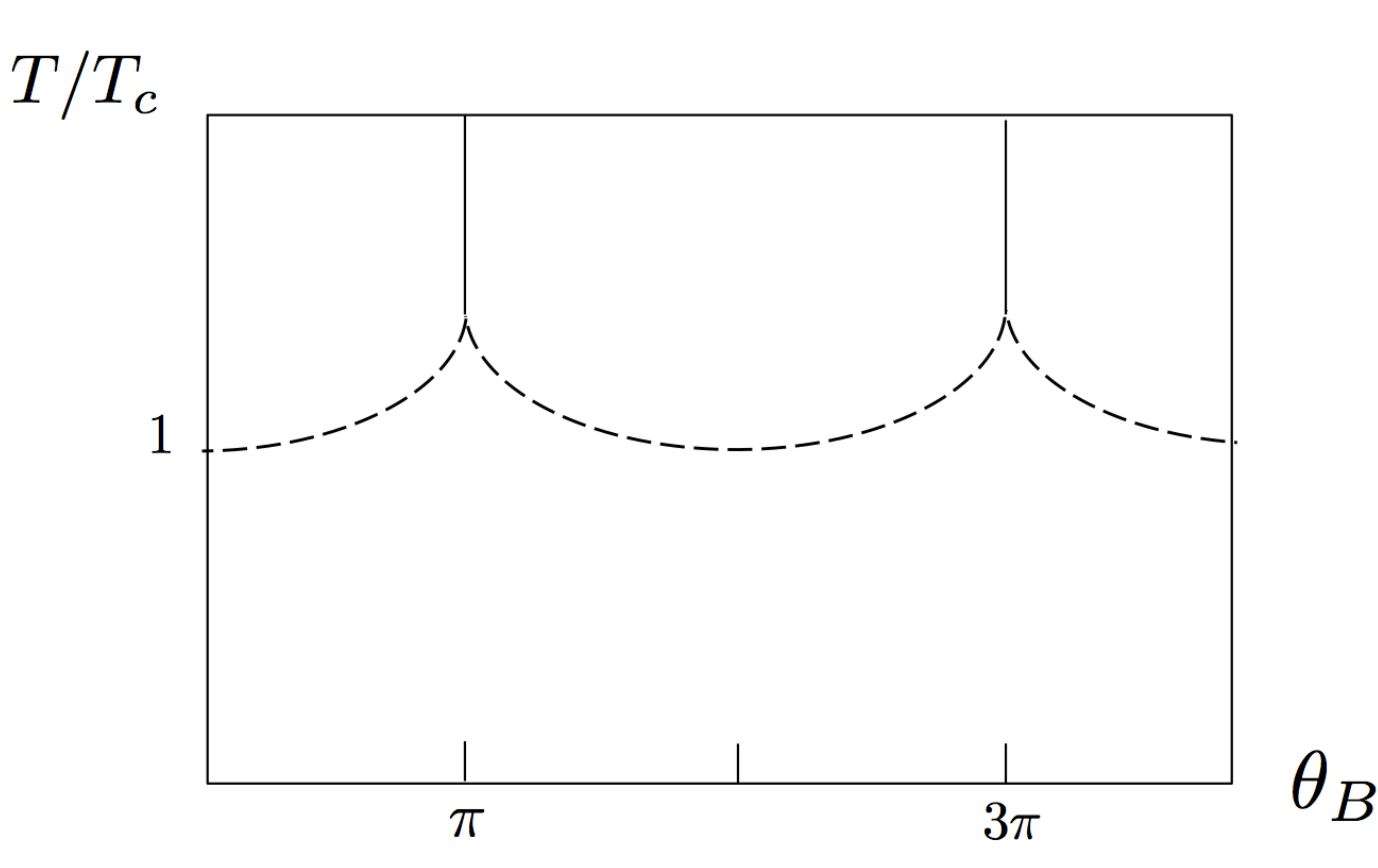}
\caption{\small{A sketch of the expected $(T,\theta)$ and $(T,\theta_B)$ phase diagrams of quenched QCD. Here $\theta$ is real and $\theta_B=\mu_B/T_c(0)$, where $\mu_B$ is the imaginary baryon chemical potential. Dashed lines are used, as in \cite{massimo2}, at the boundary between confined and deconfined phases. For the holographic model examined in this work all the lines are first order transition ones.}}
\label{fig1}
\end{figure}

As it has also been argued in recent literature (see e.g. \cite{massimo2}), at finite temperature the same structure as in (\ref{wit}) should be inherited by the Yang-Mills free energy density, provided the theory stays in the confined phase. The $(T,\theta)$ phase diagram should thus present vertical first order lines departing from the $(T=0, \theta=(2k+1)\pi)$ points. These lines are however expected to end at around $T=T_c(0)$, where a parabolic phase boundary described by an equation like (\ref{quad}) should be met. The resulting phase diagram will thus present itself as an ``arcade", with first order ``columns" emerging from $(T=0, \theta=(2k+1)\pi)$ ending at tricritical points at $T_c=T_c(\theta=(2k+1)\pi)$ where they bifurcate in two first order ``arcs". See figure \ref{fig1} for a sketch.

As observed in \cite{massimo1,massimo2}, this structure resembles that of an inverted Roberge-Weiss phase diagram \cite{RW}. 
The latter describes (quenched) QCD at finite imaginary baryon chemical potential $\mu_B$. In this case the high temperature free energy (which can be accessed perturbatively) shows discontinuities at $\theta_B = (2k+1)\pi$. 
For an $SU(N_c)$ theory with $N_f$ massless Weyl fermions in the fundamental representation, the tree level result for the $\theta_B$-dependent part of the free energy density $f (\theta_B)$ is
\be
f(\theta_B) = \frac{N_c N_f}{12} T^4 {\rm min}_k \left(\frac{\theta_B -2\pi k}{N_c}\right)^2\,.
\ee
The phase diagram (sketched in figure \ref{fig1}) is then expected to exhibit first order ``columns" emerging from the points $(T\rightarrow\infty, \theta_B = (2k+1)\pi)$. 
These lines should end on the phase boundary at $T\sim T_c(\theta_B)$ which is shown to have a quadratic dependence on $\theta_B$ on the lattice (with $T_c(\theta_B)$ increasing with $\theta_B^2$ at small $\theta_B$). The whole phase diagram looks like an inverted arcade with arcs around $T_c(0)$.

In the following we will show that the holographic QCD model studied in this paper gives a concrete realization of the structure of the above described $(T, \theta_B)$ and $(T,\theta)$ phase diagrams, which are, then, qualitatively ``dually'' related as proposed in \cite{massimo2}. 

\subsubsection{Holographic QCD at finite $T$ and $\theta_B$}

In the case of imaginary baryon chemical potential the result (\ref{tici}) readily gives
\be
\frac{T_c}{T_c(\theta_B=0)} = 1+\frac{27}{64\pi^3}N_f N_c \frac{\theta_B^2}{N_c^2}\,.
\ee
Thus, the critical temperature quadratically increases with $\theta_B$. This behavior, for the Witten-Sakai-Sugimoto model, has been already found in \cite{rafferty}. 
In the latter paper, generalizing the results of \cite{kumar}, it has also been shown that the free energy has first order discontinuities at $\theta_B = (2k+1)\pi$, so that the phase diagram is precisely as in figure \ref{fig1}. 
%The case of finite $\theta$ is just a bit more involved, as we now describe.

\subsubsection{Holographic Yang-Mills at finite $T$ and $\theta$}
We now want to construct the phase diagram of the holographic model at finite temperature and finite $\theta$-angle. 
This requires a few considerations, whose outcome is formula (\ref{tc}) - the reader uninterested in the details can skip the derivation.\footnote{See also \cite{pz} for relevant related results.}

The discussion concerns the unfavored theory dual to the backgrounds of section \ref{secreview}.
Let us take $0\le \theta < \pi$ and start by reviewing the known results at $T=0$ \cite{Witten:1998uka}, thus considering the background (\ref{witten0}). 
The $\theta$-angle is identified with the integral of a Ramond-Ramond one-form $C_1= C(u) dx_4$ along the $S_{x_4}$ circle at $u\rightarrow\infty$
\be\label{thetanorm}
\theta = \int_{S_{x_4}} C_1 = \int_{cig} F_2 \ ,
\ee
where ``$cig$'' is the $(u, x_4)$ cigar, $F_{2}=d C_1 = F_{u, x_4}du\wedge dx_4$ and we have used Stokes's theorem. 
Since the $S_{x_4}$ circle shrinks to zero size at $u=u_{0}$, regularity imposes $C(u_{0})=0$. 
With this condition, the solution of the equation of motion for $C_1$ gives
 \cite{Witten:1998uka,Barbon:2004dq}
\be
F_2 = \frac{c\, \theta}{u^4} du \wedge dx_4 \ , \qquad \Rightarrow \qquad C_1 = \left(\frac{\theta}{\bf \beta_4}-\frac{c \,\theta}{3u^3} \right) dx_{4}\,,\qquad c\equiv \frac{3 u_{0}^3}{\beta_4}\,.
\ee
Remember that $\beta_4$ is the length of the $S_{x_4}$ circle.
Under the holographic gauge/gravity map, the $\theta$-dependent contribution to the field theory vacuum energy density is mapped into the part of the gravity action (in Euclidean signature) involving $F_2$
\be
S[F_2]= -\frac{1}{4\pi(2\pi l_s)^6} \int dx^{10}\sqrt{-g} |F_2|^2\,,
\ee
which has to be computed on-shell on the solution for $F_2$ found above. Using the holographic relation 
$e^{-V_4 \varepsilon(\theta)}\approx e^{-S [F_2]}$, where $V_4$ is the (infinite) 4d Euclidean volume, one gets (see \cite{ss})
\be\label{Etheta}
\varepsilon(\theta)= \frac12 \chi_g \theta^2\,,\qquad \chi_g= \frac{1}{4(3\pi)^6}M_{KK}^4 \lambda_4^3\,.
\ee
This formula is thus of the anticipated form given in (\ref{wit}), restricted to the $0\le\theta<\pi$ range.

The $T<T_c$ geometry is precisely the same as the $T=0$ one, but for the compactification of the Euclidean time direction. 
Therefore, the holographic relation $e^{-\beta F} \approx e^{-S}$ between the free energy $F$ and the on-shell Euclidean gravity action, implies that the free energy density $f(\theta)$ in the confined phase is given by $f(\theta) = \varepsilon(\theta)$, with $\varepsilon(\theta)$ given in (\ref{Etheta}). 
In particular this means that $f(\theta)$, in this phase, is independent of $T$ and it has discontinuities at  $\theta=(2k+1)\pi$ exactly as in the $T=0$ case. 
The phase diagram at $T<T_c$ will thus show first order phase transition vertical lines emerging from the $(T=0, \theta=(2k+1)\pi)$ points.

At $T>T_c$, instead, the free energy of the model does not depend on $\theta$ (this has to be read as the leading effect in the large $N_c$ limit), since a trivial solution for $F_2$ is now admitted. 
In fact (see the background (\ref{wittenT})), the $(u, x_4)$ subspace is a now cylinder, i.e. the $S_{x_4}$ circle does not shrink anymore. 
Such a geometry allows for a trivial solution $C(u)\sim\theta$, i.e. $F_2=0$ of the equation of motion of $C_1$. 
This is actually the solution which minimizes the gravity action. 
The latter (which depends on $F_2$) will thus be independent of $\theta$. 
As a result, the field theory free energy will not depend on the $\theta$ angle. 
We now show that this implies a structure of the form given in (\ref{quad}) for the $\theta$-dependence of the deconfining temperature. 

When $\theta=0$, the difference between the free energy density at $T<T_c$ and that at $T>T_c$ in the unflavored case %, can be read from formula (39) in the Appendix of \cite{Aharony:2006da}.
is given by the difference of (\ref{gibbs}) and (\ref{pconf}) with $\epsilon_f=\epsilon_{f\,T}=0$ (see \cite{Aharony:2006da}\footnote{A typo in the overall coefficient in \cite{Aharony:2006da} has been corrected here.}).
If we want to compute the same quantity at finite $\theta$ it just suffices to add the $T<T_c$ contribution $f(\theta)=\varepsilon(\theta)$ as given in (\ref{Etheta}). The result is 
\be
\Delta f(\theta) = \Delta f(\theta=0) - \varepsilon(\theta) = -  \left(\frac{2 N_c^2 \lambda_4}{2187 \pi^2 M_{KK}^2}\right) \left[(2\pi T)^6 - M_{KK}^6 + \kappa M_{KK}^6\lambda_4^2\frac{\theta^2}{N_c^2}\right]\,,
\ee
where
\be
\kappa \equiv \frac{3}{16 \cdot \pi^{4}}\ .
\label{Kappa}
\ee
Thus, the critical temperature for the phase transition at fixed $\theta$ is found to be given by
\be\label{tc}
\frac{T_c(\theta)}{T_{c}(0)} = \left(1 - \kappa \lambda_4^2\frac{\theta^2}{N_c^2} \right)^{1/6}\approx 1 - \frac{\kappa}{6}\lambda_4^2\frac{\theta^2}{N_c^2}\ ,
\ee
where $T_{c}(0)=M_{KK}/2\pi$ is the value at $\theta=0$, and we have expanded the parenthesis considering that the second term is suppressed w.r.t. 1 for small $\theta/N_c$.

Thus, $T_c$ is, consistently with large $N_c$ estimates, a quadratically decreasing function of $\theta$ as in (\ref{quad}) with coefficient
\be
r_{\theta} = \frac{\kappa}{6} \lambda_4^2\,.
\ee
The numerical value of $r_{\theta}$ is of the same order of magnitude of the one extracted from lattice QCD or large-$N_c$ considerations \cite{massimo1}, provided $\lambda_4 \sim 25$. The latter is close to the value commonly used to compare results from the WSS model to phenomenology.
The periodicity of the physics with $\theta$ dictates the whole phase diagram to have the periodic structure of an inverted Roberge-Weiss \cite{RW} one (as in figure \ref{fig1}), with cusps joining the first-order lines emanating from $T=0$ at $\theta = (2n+1)\pi$. 

All in all, the structure of the phase diagram of the holographic model is the same one advocated in lattice QCD (fig. \ref{fig1}), both at finite $\theta$ and imaginary chemical potential.
The qualitative ``duality''\footnote{Needless to say, this is not a precise duality but only an interesting qualitative analogy.} of the two phase diagrams in the holographic model depends crucially on the symmetry of the $T<T_c$ and $T>T_c$ backgrounds (\ref{witten0}), (\ref{wittenT}) under the exchange of the $S_{x_0}$ and $S_{x_4}$ circles.
In fact, the latter determine the behavior of the free energy density with $\theta$ and $\theta_B$, as they support the gravity one-forms $C_1$ and ${\cal A}_1$ dual to these quantities.
The quadratic dependence of $f$ on $\theta$ and $\theta_B$ just follows from the Maxwell form of the action for the one-form fields.\footnote{This is true for ${\cal A}_1$ in the small charge regime we are considering. At larger charges the non linear terms in ${\cal A}_1'^2$ in the Born-Infeld action could correct this behavior.}
As usual, the holographic description nicely geometrizes properties of the dual field theory.

\subsection{Degrees of freedom for the phase diagrams}
In the stringy picture there are degrees of freedom which emerge naturally in the confined and deconfined phases of the theory.
In fact, $D0$-branes are electrically charged under $C_1$ (related to the $\theta$ angle). 
This provides a way for identifying the instantons in the model as Euclidean $D0$-branes wrapped along the $S_{x_4}$ circle.\footnote{See e.g. \cite{Bergman:2006xn,pz}
 and references therein.} Since at $T> T_c$ the circle does not shrink, this configuration is stable and maps to a instanton gas configuration in the dual field theory. 
At $T<T_c$ this picture does not hold anymore and this is realized by the fact that the wrapped $D0$-brane is now shrinking with the cycle. A natural question to ask, thus, is what replaces the instantons in the confined phase. A possible hint, as suggested in \cite{massimo1}, could come from the comparison with what happens in the flavored case for the ``dual'' phase diagram at imaginary baryon chemical potential.

The degrees of freedom replacing the instantons in the confined phase are baryons (recall that in the WSS model, a baryon vertex, which is identified with a $\tilde D0$-brane obtained by wrapping $D4$-branes on the transverse $S^4$, is electrically charged under the $U(1)_B$ gauge field ${\cal A}_t$). 
In the deconfined phase these are replaced by quarks which have fractional baryon charge. This, could suggest that in the ``dual'' $(T,\theta)$ setup, the degrees of freedom replacing the instantons in the confined phase are actually ``fractional instantons''. Understanding their role in the present setup as well as in real world Yang-Mills theory could possibly uncover relevant aspects of the dynamics of confinement. 
%%%%%%%%%%%%%%%%%%%%%%%%%%%%%%%%%%%%%%%%%%%%%%%%%
\section{Conclusions and discussion}\label{secconclusions}

In this paper we have constructed gravity backgrounds corresponding to the Witten-Sakai-Sugimoto (WSS) model with backreaction of the $D8$-branes, that is with dual dynamical flavors.
We have provided solutions at small (or zero) and large temperature and at finite chemical potential (or charge density).
The solutions are suitable for the study of the influence of dynamical flavors on physical observables in a setting (the WSS theory) which represents the top-down model closest to (planar) QCD.

In particular, in this paper we have concentrated our attention on the phase diagrams, re-deriving the thermodynamic of the system and discussing it with an eye to recent lattice QCD literature.
We have observed qualitative similarities between the QCD and WSS theories in the finite temperature, finite baryon chemical potential regimes. 
The similarities extend to the finite $\theta$ angle and finite imaginary chemical potential regimes too.
As usual, the holographic approach, while describing a theory which is not exactly (planar) QCD, has the virtue of geometrizing certain aspects of the physics and suggesting information on the possible relevant degrees of freedom. 

Moreover, we have provided covariant counterterms for the $D8$-branes in the Witten-Sakai-Sugimoto model allowing for its standard holographic renormalization.
As far as we know, these counterterms have never been explicitly reported in the vast literature on the model.
Our main motivation for deriving these counterterms has been to holographically renormalize the free energies, in order to study the phase diagrams and the confinement/deconfinement transition.
In many examples of holographic theories (see e.g. \cite{Cotrone:2007qa,Bigazzi:2009bk} for flavored cases) holographic renormalization is not needed, since the background subtraction method gives a finite result for the difference of free energies.
Interestingly, this is not the case for the WSS model.
In fact, the $D8$-branes have different sub-leading divergences in the $T<T_c$ and $T>T_c$ phases, so they need to be renormalized independently (this is the reason for having different coefficients of the counterterms in the two cases).
Note also that these subleading divergences carry information on the IR region of the theory (they depend on $u_0$ or $u_T$), suggesting that there could be no clear-cut separation of the UV and IR regimes.

Anyway, despite the presence of a Landau pole due to irrelevant operators, the theory admits a standard holographic renormalization (to be possibly improved in the 1d hamiltonian formalism of \cite{papad}, as already commented).
This might depend on the fact that we have worked at leading order in the flavor counting parameter $\epsilon_f \sim \lambda_4 N_f/N_c$, as advocated in \cite{vanRees:2011fr}.
The situation might be more involved at higher orders.
Moreover, while we have renormalized the gravity action, relevant for the computation of correlators of local operators, a separate renormalization prescription is needed for non-local observables, such as Wilson loops.
We plan to come back to this interesting issue in the future.

As said, the solutions we have constructed in this paper include the backreaction of the $D8$-branes at leading order in the flavor counting parameter $\epsilon_f$.
This allows to find analytic solutions.
Going beyond the leading order would face two type of problems.
The first one is technical: finding analytic solution at higher orders appears to be a hopeless task and one should resort to a numeric investigation, which is not guaranteed to be straightforward.
In fact, higher order terms in $\epsilon_f$ require to add gravity fields in the setting, especially in the charged case (as in the $D3-D7$ system \cite{Bigazzi:2009bk,Bigazzi:2011it,Magana:2012kh,Ammon:2012qs,Cotrone:2012um,Bigazzi:2013jqa}).
This causes the proliferation of degrees of freedom and a complication of the equations to be solved.
Already finding analytic IR asymptotics can be problematic in this situation.
Moreover, the presence of the UV Landau pole demands to solve the equations up to a finite cut-off, which brings in the game a certain degree of uncertainty and technical complications.
Finally, at higher order in $\epsilon_f$, there could be corrections to the DBI action, such as blackfold terms, taking into account thermalization of the brane degrees of freedom \cite{Emparan:2009cs};
% ,Emparan:2009at,Grignani:2010xm,Grignani:2013ewa
it would be extremely difficult to include such contributions in the WSS theory.

On top of this technical complications, there is a second type of problem, having to do with the energy scale of the Landau pole.
As one can readily realize, and as can be seen from (\ref{LPposition}), the energy scale of the Landau pole is inversely proportional to $\epsilon_f$.
Requiring a finite separation between the Landau pole scale and the IR physics, forces us to stay in the small $\epsilon_f$ regime.
Thus, the leading $\epsilon_f$-solutions we have constructed furnish the main perturbative contribution to the exact solution.\footnote{Possible non-perturbative $\epsilon_f$-effects would be a very interesting topic.}
In this sense, knowing the exact solution is not a urgent priority.

Since the small $\epsilon_f$ regime is mandatory in this model, one could think that the probe approximation is sufficient to capture all the relevant physics.
This is only partially true: the probe and backreaction approaches are more complementary than equivalent.
In fact, whenever calculations can be performed both in the probe approximation and in the leading $\epsilon_f$ case, the results should coincide.
Thus, one would likely prefer to use the easier approach of the probe approximation.
On the other hand, for many observables one does not have a given prescription on how to extract information from the probe - this has to be derived case by case.
Instead, with the backreacted solution one can use the standard prescriptions derived for the unflavored theory, involving only the background, to compute observables in a straightforward way (we have given examples in section \ref{secphys}).

Coming to our other approximation, it would be certainly useful and worthwhile to try and solve equations (\ref{eqsq}) beyond the small charge regime.
There might exist analytic solutions, which would allow to study the physics in a regime where lattice QCD simulations are problematic.
Other obvious venues for future research can include, among the others, the study of transport properties of the theory (as done in \cite{Bigazzi:2009tc}
%,Bigazzi:2010ku,Tarrio:2013tta} 
for the $D3-D7$ system) and the flavor and charge dependence of the entanglement entropy \cite{Kim:2013ysa}.
%,Chang:2013mca,Kontoudi:2013rla,Karch:2014ufa,Kol:2014nqa,Chang:2014oia}.

Moreover, it would be interesting to consider the backreaction of explicit baryonic sources, so to extend the results of, e.g. \cite{bergman,Kumar:2012ui}.
%,Faedo:2013aoa,Bolognesi:2014dja}.}

%%%%%%%%%%%%%%%%%%%%%%%%%%%%%%%%%%%%%%%%%%%%%%%%
\vskip 15pt \centerline{\bf Acknowledgments} \vskip 10pt \noindent We are grateful to Alfonso Ballon-Bayona, Paolo Benincasa, Stefano Bolognesi, Claudio Bonati, Massimo D'Elia, Maria Paola Lombardo, Ioannis Papadimitriou, Domenico Seminara, Kostas Skenderis, Shigeki Sugimoto and Ettore Vicari for relevant comments and discussions. 

%%%%%%%%%%%%%%%%%%%%%%%%%%%%%%%%%%%%%%%
\appendix
\section{The $u_0=u_T=0$ limit}
Let us consider the limiting case in which both the $x_4$ and the temporal circles do not shrink. This is like starting with the flat $D4$-brane background and trivially ``thermalizing" it, taking the $x_0$ and $x_4$ coordinates to be compactified on circles of length $1/T$ and $2\pi/M_{KK}$ respectively.

The addition of smeared $D8$ flavor branes as in the main body of this work leads to a background with string frame metric
\be
ds^2 = e^{2\lambda}\left[dx_{\mu}dx^{\mu} + e^{2(\tilde\lambda-\lambda)}dx_4^2\right] + l_s^2 e^{-2\varphi}d\rho^2 + l_s^2 e^{2\nu}d\Omega_4^2\,,
\ee
where $\varphi=2\phi-4\lambda-2\tilde\lambda-4\nu$. In the absence of backreaction we have
\bea
&&\lambda_0= \tilde\lambda_0 = \frac14 \log\left[\frac{l_s^3}{R^3}\frac{g_s^2}{3\rho}\right]\,,\nonumber\\
&&\phi_0 = \lambda_0 + \log(g_s)\,,\nonumber\\
&&\nu_0 = \frac13\lambda_0 + \log \frac{R}{l_s}\,.
\eea
Defining
\be
u_s\equiv l_s g_s^{2/3}\,, \qquad \epsilon_{f\,s} \equiv \frac{R^{3/2} u_{s}^{1/2} g_s}{l_s^2} Q_f = \epsilon_f\sqrt{\frac{u_s}{u_0}}\,,
\ee
and expanding the various functions in $\epsilon_{f\,s}$, we get for the first order corrections
\bea
&&\lambda_1 = \frac{101}{455 (3\rho)^{1/6}}+\gamma_1\rho^2+\frac{\gamma_2}{\rho}+\frac14 (\alpha_1-\alpha_2)+\frac14(\beta_1-\beta_2)\rho\,,\nonumber \\
&& \tilde\lambda_1 = \lambda_1 - \frac{2}{7(3\rho)^{1/6}}-\alpha_1-\beta_1\rho\,,\nonumber \\
&& \phi_1 = \lambda_1 +\frac{4}{7(3\rho)^{1/6}}-\alpha_2-\beta_2\rho\,,\nonumber \\
&& \nu_1 = \frac{1}{12}(\alpha_1-5\alpha_2+(\beta_1-5\beta_2)\rho )+\frac{25}{91(3\rho)^{1/6}}+ \rho^2\mu_1 + \frac{\mu_2}{\rho}\,.
\eea
Requiring regularity in the IR (i.e. at $\rho\rightarrow\infty$) and turning off the leading UV divergences (i.e. those at $\rho\rightarrow0$) amounts on setting
\be
\gamma_1=\beta_1=\beta_2=\mu_1=0\,,\qquad \gamma_2=\mu_2=0\,.
\ee
Imposing these conditions we get the simple solution
\bea 
&&\tilde\lambda_1 -\lambda_1 = -\alpha_1 -\frac{2}{7(3\rho)^{1/6}}\,,\nonumber \\
&& \phi_1-\lambda_1 = -\alpha_2 + \frac{4}{7(3\rho)^{1/6}}\,,\nonumber \\
&& \phi_1 - 3\nu_1=-\frac{2}{65 (3\rho)^{1/6}}\,,\nonumber \\
&& \phi_1 + 2\tilde\lambda_1 -3\lambda_1 = -2\alpha_1 -\alpha_2\,.
\eea
%%%%%%%%%%%%%%%%%%%%%%%%%%%%%%%%%%%%%%
%%%%%%%%%%%%%%%%%%%%%%%%%%%%%%%%%%%%%%
  

\begin{thebibliography}{99} 
\bibitem{witten} E.~Witten,
  ``Anti-de Sitter space, thermal phase transition, and confinement in gauge theories,''
  Adv.\ Theor.\ Math.\ Phys.\  {\bf 2}, 505 (1998)
  [hep-th/9803131].
  %%CITATION = HEP-TH/9803131;%%  
\bibitem{ss} T.~Sakai and S.~Sugimoto, ``Low energy hadron physics in holographic QCD,''
  Prog.\ Theor.\ Phys.\  {\bf 113}, 843 (2005)
  [hep-th/0412141].
  %%CITATION = HEP-TH/0412141;%% 
\bibitem{Aharony:2006da} 
  O.~Aharony, J.~Sonnenschein and S.~Yankielowicz,
  ``A Holographic model of deconfinement and chiral symmetry restoration,''
  Annals Phys.  {\bf 322}, 1420 (2007)
  [hep-th/0604161].
  %%CITATION = HEP-TH/0604161;%% 
\bibitem{Kim:2006gp} 
  K.~Y.~Kim, S.~J.~Sin and I.~Zahed,
  ``Dense hadronic matter in holographic QCD,''
  J.\ Korean Phys.\ Soc.\  {\bf 63}, 1515 (2013)
  [hep-th/0608046].
  %%CITATION = HEP-TH/0608046;%%
  N.~Horigome and Y.~Tanii,
  ``Holographic chiral phase transition with chemical potential,''
  JHEP {\bf 0701}, 072 (2007)
  [hep-th/0608198].
  %%CITATION = HEP-TH/0608198;%%
 \bibitem{bergman}  D.~K.~Hong, M.~Rho, H.~U.~Yee and P.~Yi,
  ``Chiral Dynamics of Baryons from String Theory,''
  Phys.\ Rev.\ D {\bf 76}, 061901 (2007)
  [hep-th/0701276].
  %%CITATION = HEP-TH/0701276;%%
  H.~Hata, T.~Sakai, S.~Sugimoto and S.~Yamato,
  ``Baryons from instantons in holographic QCD,''
  Prog.\ Theor.\ Phys.\  {\bf 117}, 1157 (2007)
  [hep-th/0701280].
  %%CITATION = HEP-TH/0701280;%%
  O.~Bergman, G.~Lifschytz and M.~Lippert, ``Holographic Nuclear Physics,''
  JHEP {\bf 0711}, 056 (2007)
  [arXiv:0708.0326 [hep-th]].
  %%CITATION = ARXIV:0708.0326;%% 
 \bibitem{massimo1}M.~D'Elia and F.~Negro,
  ``$\theta$ dependence of the deconfinement temperature in Yang-Mills theories,''
  Phys.\ Rev.\ Lett.\  {\bf 109}, 072001 (2012)
  [arXiv:1205.0538 [hep-lat]].
  %%CITATION = ARXIV:1205.0538;%%
\bibitem{Bigazzi:2005md} 
  F.~Bigazzi, R.~Casero, A.~L.~Cotrone, E.~Kiritsis and A.~Paredes,
  ``Non-critical holography and four-dimensional CFT's with fundamentals,''
  JHEP {\bf 0510}, 012 (2005)
  [hep-th/0505140].
  %%CITATION = HEP-TH/0505140;%%
 %\cite{Bigazzi:2011db}
%\cite{Casero:2006pt}
\bibitem{Casero:2006pt} 
  R.~Casero, C.~Nunez and A.~Paredes,
  ``Towards the string dual of N=1 SQCD-like theories,''
  Phys.\ Rev.\ D {\bf 73}, 086005 (2006)
  [hep-th/0602027].
  %%CITATION = HEP-TH/0602027;%%  
\bibitem{Benini:2006hh} 
  F.~Benini, F.~Canoura, S.~Cremonesi, C.~Nunez and A.~V.~Ramallo,
  ``Unquenched flavors in the Klebanov-Witten model,''
  JHEP {\bf 0702}, 090 (2007)
  [hep-th/0612118].
  %%CITATION = HEP-TH/0612118;%%
\bibitem{Nunez:2010sf} 
  C.~Nunez, A.~Paredes and A.~V.~Ramallo,
  ``Unquenched Flavor in the Gauge/Gravity Correspondence,''
  Adv.\ High Energy Phys.\  {\bf 2010}, 196714 (2010)
  [arXiv:1002.1088 [hep-th]].
  %%CITATION = ARXIV:1002.1088;%% 
\bibitem{Bigazzi:2011db} 
  F.~Bigazzi, A.~L.~Cotrone, J.~Mas, D.~Mayerson and J.~Tarrio,
  ``Holographic Duals of Quark Gluon Plasmas with Unquenched Flavors,''
  Commun.\ Theor.\ Phys.\  {\bf 57}, 364 (2012)
  [arXiv:1110.1744 [hep-th]].
  %%CITATION = ARXIV:1110.1744;%%   
\bibitem{sonne} B.~A.~Burrington, V.~S.~Kaplunovsky and J.~Sonnenschein,
  ``Localized Backreacted Flavor Branes in Holographic QCD,''
  JHEP {\bf 0802}, 001 (2008)
  [arXiv:0708.1234 [hep-th]].
  %%CITATION = ARXIV:0708.1234;%%
  \bibitem{Antonyan:2006vw} 
  E.~Antonyan, J.~A.~Harvey, S.~Jensen and D.~Kutasov,
  ``NJL and QCD from string theory,''
  hep-th/0604017.
  %%CITATION = HEP-TH/0604017;%%    
\bibitem{Mandal:2011ws} 
  G.~Mandal and T.~Morita,
  ``Gregory-Laflamme as the confinement/deconfinement transition in holographic QCD,''
  JHEP {\bf 1109}, 073 (2011)
  [arXiv:1107.4048 [hep-th]].
  %%CITATION = ARXIV:1107.4048;%%  

\bibitem{Bigazzi:2011it} 
  F.~Bigazzi, A.~L.~Cotrone, J.~Mas, D.~Mayerson and J.~Tarrio,
  ``D3-D7 Quark-Gluon Plasmas at Finite Baryon Density,''
  JHEP {\bf 1104}, 060 (2011)
  [arXiv:1101.3560 [hep-th]].
  %%CITATION = ARXIV:1101.3560;%%
\bibitem{Cotrone:2012um} 
  A.~L.~Cotrone and J.~Tarrio,
  ``Consistent reduction of charged D3-D7 systems,''
  JHEP {\bf 1210}, 164 (2012)
  [arXiv:1207.6703 [hep-th]].
  %%CITATION = ARXIV:1207.6703;%% 
%\cite{Bigazzi:2013jqa}
\bibitem{Bigazzi:2013jqa} 
  F.~Bigazzi, A.~L.~Cotrone and J.~Tarrio,
  ``Charged D3-D7 plasmas: novel solutions, extremality and stability issues,''
  JHEP {\bf 1307}, 074 (2013)
  [arXiv:1304.4802 [hep-th]].
  %%CITATION = ARXIV:1304.4802;%%

\bibitem{Ooguri:2010xs} 
  H.~Ooguri and C.~S.~Park,
  ``Spatially Modulated Phase in Holographic Quark-Gluon Plasma,''
  Phys.\ Rev.\ Lett.\  {\bf 106}, 061601 (2011)
  [arXiv:1011.4144 [hep-th]].
  %%CITATION = ARXIV:1011.4144;%%  
 \bibitem{noiWL}  F.~Bigazzi, A.~L.~Cotrone, L.~Martucci and L.~A.~Pando Zayas,
  ``Wilson loop, Regge trajectory and hadron masses in a Yang-Mills theory from semiclassical strings,''
  Phys.\ Rev.\ D {\bf 71}, 066002 (2005)
  [hep-th/0409205].
  %%CITATION = HEP-TH/0409205;%% 
\bibitem{giata}  D.~Giataganas and N.~Irges,
  ``Flavor Corrections in the Static Potential in Holographic QCD,''
  Phys.\ Rev.\ D {\bf 85}, 046001 (2012)
  [arXiv:1104.1623 [hep-th]].
  %%CITATION = ARXIV:1104.1623;%%
\bibitem{Witten:1998xy} 
  E.~Witten,
 ``Baryons and branes in anti-de Sitter space,''
  JHEP {\bf 9807}, 006 (1998)
  [hep-th/9805112].
  %%CITATION = HEP-TH/9805112;%%\end{document} 
\bibitem{skenderis} I.~Kanitscheider, K.~Skenderis and M.~Taylor,
  ``Precision holography for non-conformal branes,''
  JHEP {\bf 0809}, 094 (2008)
  [arXiv:0807.3324 [hep-th]].
  %%CITATION = ARXIV:0807.3324;%%    
\bibitem{Bigazzi:2009gu} 
  F.~Bigazzi, A.~L.~Cotrone, A.~Paredes and A.~V.~Ramallo,
  ``Screening effects on meson masses from holography,''
  JHEP {\bf 0905}, 034 (2009)
  [arXiv:0903.4747 [hep-th]].
  %%CITATION = ARXIV:0903.4747;%%  
\bibitem{russo} J.~G.~Russo and K.~Sfetsos,
  ``Rotating D3-branes and QCD in three-dimensions,''
  Adv.\ Theor.\ Math.\ Phys.\  {\bf 3}, 131 (1999)
  [hep-th/9901056].
  %%CITATION = HEP-TH/9901056;%%  
\bibitem{ballon} A.~Ballon-Bayona,
  ``Holographic deconfinement transition in the presence of a magnetic field,''
  JHEP {\bf 1311}, 168 (2013)
  [arXiv:1307.6498 [hep-th]].
  %%CITATION = ARXIV:1307.6498;%%
\bibitem{myersmateos} D.~Mateos, R.~C.~Myers and R.~M.~Thomson, ``Thermodynamics of the brane,'' JHEP {\bf 0705}, 067 (2007) [hep-th/0701132].
  %%CITATION = HEP-TH/0701132;%%
\bibitem{Benincasa:2009ze}  P.~Benincasa, ``A note on Holographic Renormalization of Probe D-Branes,''
arXiv:0903.4356 [hep-th].
  %%CITATION = ARXIV:0903.4356;%%  
\bibitem{papad}  I.~Papadimitriou,
 ``Holographic renormalization as a canonical transformation,''
  JHEP {\bf 1011}, 014 (2010)
  [arXiv:1007.4592 [hep-th]];
  %%CITATION = ARXIV:1007.4592;%%  I.~Papadimitriou,
 ``Holographic Renormalization of general dilaton-axion gravity,''
  JHEP {\bf 1108}, 119 (2011)
  [arXiv:1106.4826 [hep-th]].
  %%CITATION = ARXIV:1106.4826;%%
\bibitem{ihl}  M.~Ihl, A.~Kundu and S.~Kundu,
  ``Back-reaction of Non-supersymmetric Probes: Phase Transition and Stability,''
  JHEP {\bf 1212}, 070 (2012)
  [arXiv:1208.2663 [hep-th]].
  %%CITATION = ARXIV:1208.2663;%%
 M.~Sohaib Alam, M.~Ihl, A.~Kundu and S.~Kundu,
 ``Dynamics of Non-supersymmetric Flavours,''
  JHEP {\bf 1309}, 130 (2013)
  [arXiv:1306.2178 [hep-th]].
  %%CITATION = ARXIV:1306.2178;%% 
\bibitem{parnakim} M.~Kulaxizi and A.~Parnachev,
  ``Holographic Responses of Fermion Matter,''
  Nucl.\ Phys.\ B {\bf 815}, 125 (2009)
  [arXiv:0811.2262 [hep-th]].
  %%CITATION = ARXIV:0811.2262;%%  
 K.~y.~Kim and J.~Liao,
 ``On the Baryonic Density and Susceptibilities in a Holographic Model of QCD,''
  Nucl.\ Phys.\ B {\bf 822}, 201 (2009)
  [arXiv:0906.2978 [hep-th]].
  %%CITATION = ARXIV:0906.2978;%% 
\bibitem{mpl}A.~Deuzeman, M.~P.~Lombardo, K.~Miura, T.~Nunes da Silva and E.~Pallante,
  `Phases of many flavors QCD : Lattice results,''
  PoS ConfinementX {\bf }, 274 (2012)
  [arXiv:1304.3245 [hep-lat]].
  %%CITATION = ARXIV:1304.3245;%%
 \bibitem{qcddari} J.~Braun and H.~Gies,
  ``Scaling laws near the conformal window of many-flavor QCD,''
  JHEP {\bf 1005}, 060 (2010)
  [arXiv:0912.4168 [hep-ph]].
  %%CITATION = ARXIV:0912.4168;%% 
 \bibitem{shuryak}
J.~Liao and E.~Shuryak,
  ``Effect of Light Fermions on the Confinement Transition in QCD-like Theories,''
  Phys.\ Rev.\ Lett.\  {\bf 109}, 152001 (2012)
  [arXiv:1206.3989 [hep-ph]].
  %%CITATION = ARXIV:1206.3989;%%
\bibitem{Bigazzi:2009bk} 
  F.~Bigazzi, A.~L.~Cotrone, J.~Mas, A.~Paredes, A.~V.~Ramallo and J.~Tarrio,
  ``D3-D7 Quark-Gluon Plasmas,''
  JHEP {\bf 0911}, 117 (2009)
  [arXiv:0909.2865 [hep-th]].
  %%CITATION = ARXIV:0909.2865;%%  
 \bibitem{massimo2}M.~D'Elia and F.~Negro,
 ``Phase diagram of Yang-Mills theories in the presence of a $\theta$ term,''
  Phys.\ Rev.\ D {\bf 88}, no. 3, 034503 (2013)
  [arXiv:1306.2919 [hep-lat]].
  %%CITATION = ARXIV:1306.2919;%%
  \bibitem{vicari} E.~Vicari and H.~Panagopoulos,
  ``Theta dependence of SU(N) gauge theories in the presence of a topological term,''
  Phys.\ Rept.\  {\bf 470}, 93 (2009)
  [arXiv:0803.1593 [hep-th]].
  %%CITATION = ARXIV:0803.1593;%% \end{thebibliography}
\bibitem{imchem}M.~G.~Alford, A.~Kapustin and F.~Wilczek,
  ``Imaginary chemical potential and finite fermion density on the lattice,''
  Phys.\ Rev.\ D {\bf 59}, 054502 (1999)
  [hep-lat/9807039].
  %%CITATION = HEP-LAT/9807039;%%
 A.~Hart, M.~Laine and O.~Philipsen,
 ``Testing imaginary versus real chemical potential in finite temperature QCD,''
  Phys.\ Lett.\ B {\bf 505}, 141 (2001)
  [hep-lat/0010008].
  %%CITATION = HEP-LAT/0010008;%%
P.~de Forcrand and O.~Philipsen,
 ``The QCD phase diagram for small densities from imaginary chemical potential,''
  Nucl.\ Phys.\ B {\bf 642}, 290 (2002)
  [hep-lat/0205016].
  %%CITATION = HEP-LAT/0205016;%%
 M.~D'Elia and M.~-P.~Lombardo,
 ``Finite density QCD via imaginary chemical potential,''
  Phys.\ Rev.\ D {\bf 67}, 014505 (2003)
  [hep-lat/0209146]. 
%%CITATION = HEP-LAT/0209146;%%
\bibitem{imtheta}V.~Azcoiti, G.~Di Carlo, A.~Galante and V.~Laliena,
  ``New proposal for numerical simulations of theta vacuum - like systems,''
  Phys.\ Rev.\ Lett.\  {\bf 89}, 141601 (2002)
  [hep-lat/0203017].
  %%CITATION = HEP-LAT/0203017;%%. 
  B.~Alles and A.~Papa,
 ``Mass gap in the 2D O(3) non-linear sigma model with a theta=pi term,''
  Phys.\ Rev.\ D {\bf 77}, 056008 (2008)
  [arXiv:0711.1496 [cond-mat.stat-mech]].
  %%CITATION = ARXIV:0711.1496;%%
S.~Aoki, R.~Horsley, T.~Izubuchi, Y.~Nakamura, D.~Pleiter, P.~E.~L.~Rakow, G.~Schierholz and J.~Zanotti,
 ``The Electric dipole moment of the nucleon from simulations at imaginary vacuum angle theta,''
  arXiv:0808.1428 [hep-lat].
  %%CITATION = ARXIV:0808.1428;%%  
 H.~Panagopoulos and E.~Vicari,
 ``The 4D SU(3) gauge theory with an imaginary $\theta$ term,''
  JHEP {\bf 1111}, 119 (2011)
  [arXiv:1109.6815 [hep-lat]].
  %%CITATION = ARXIV:1109.6815;%% 
  C.~Bonati, M.~D'Elia, H.~Panagopoulos and E.~Vicari,
 ``Change of theta dependence in 4D SU(N) gauge theories across the deconfinement transition,''
  Phys.\ Rev.\ Lett.\  {\bf 110}, 252003 (2013)
  [arXiv:1301.7640 [hep-lat]].
  %%CITATION = ARXIV:1301.7640;%%   
\bibitem{wittold} E.~Witten,
  ``Large N Chiral Dynamics,''
  Annals Phys.\  {\bf 128}, 363 (1980).
  %%CITATION = APNYA,128,363;%%
%\cite{Witten:1998uka}
\bibitem{Witten:1998uka} 
  E.~Witten,
  ``Theta dependence in the large N limit of four-dimensional gauge theories,''
  Phys.\ Rev.\ Lett.\  {\bf 81}, 2862 (1998)
  [hep-th/9807109].
  %%CITATION = HEP-TH/9807109;%%
  %152 citations counted in INSPIRE as of 25 Jun 2013 
\bibitem{RW}A.~Roberge and N.~Weiss,
  ``Gauge Theories With Imaginary Chemical Potential And The Phases Of Qcd,''
  Nucl.\ Phys.\ B {\bf 275}, 734 (1986).
  %%CITATION = NUPHA,B275,734;%%
  %190 citations counted in INSPIRE as of 16 Jul 2013  
\bibitem{rafferty}  J.~Rafferty,
 ``Holographic Roberge Weiss Transitions II - Defect Theories and the Sakai Sugimoto Model,''
  JHEP {\bf 1109}, 087 (2011)
  [arXiv:1103.2315 [hep-th]].
  %%CITATION = ARXIV:1103.2315;%% 
\bibitem{kumar}G.~Aarts, S.~P.~Kumar and J.~Rafferty,
 ``Holographic Roberge-Weiss Transitions,''
  JHEP {\bf 1007}, 056 (2010)
  [arXiv:1005.2947 [hep-th]].
  %%CITATION = ARXIV:1005.2947;%%%\\
%\cite{Bergman:2006xn}
\bibitem{Bergman:2006xn} 
O.~Bergman and G.~Lifschytz,
``Holographic U(1)(A) and String Creation,''
JHEP {\bf 0704}, 043 (2007)
[hep-th/0612289].
\bibitem{pz}A.~Parnachev and A.~R.~Zhitnitsky,
  ``Phase Transitions, theta Behavior and Instantons in QCD and its Holographic Model,''
  Phys.\ Rev.\ D {\bf 78}, 125002 (2008)
  [arXiv:0806.1736 [hep-ph]].
  %%CITATION = ARXIV:0806.1736;%%  
  A.~S.~Gorsky, V.~I.~Zakharov and A.~R.~Zhitnitsky,
  ``On Classification of QCD defects via holography,''
  Phys.\ Rev.\ D {\bf 79}, 106003 (2009)
  [arXiv:0902.1842 [hep-ph]].
\bibitem{Barbon:2004dq} 
  J.~L.~F.~Barbon, C.~Hoyos-Badajoz, D.~Mateos and R.~C.~Myers,
  ``The Holographic life of the eta-prime,''
  JHEP {\bf 0410}, 029 (2004)
  [hep-th/0404260].
  %%CITATION = HEP-TH/0404260;%%
\bibitem{Cotrone:2007qa} 
  A.~L.~Cotrone, J.~M.~Pons and P.~Talavera,
  ``Notes on a SQCD-like plasma dual and holographic renormalization,''
  JHEP {\bf 0711}, 034 (2007)
  [arXiv:0706.2766 [hep-th]].
  %%CITATION = ARXIV:0706.2766;%%
\bibitem{vanRees:2011fr} 
  B.~C.~van Rees,
  ``Holographic renormalization for irrelevant operators and multi-trace counterterms,''
  JHEP {\bf 1108}, 093 (2011)
  [arXiv:1102.2239 [hep-th]].
  %%CITATION = ARXIV:1102.2239;%%
\bibitem{Magana:2012kh} 
  A.~Magana, J.~Mas, L.~Mazzanti and J.~Tarrio,
  ``Probes on D3-D7 Quark-Gluon Plasmas,''
  JHEP {\bf 1207}, 058 (2012)
  [arXiv:1205.6176 [hep-th]].
  %%CITATION = ARXIV:1205.6176;%% 
\bibitem{Ammon:2012qs} 
  M.~Ammon, V.~G.~Filev, J.~Tarrio and D.~Zoakos,
  ``D3/D7 Quark-Gluon Plasma with Magnetically Induced Anisotropy,''
  JHEP {\bf 1209}, 039 (2012)
  [arXiv:1207.1047 [hep-th]].
  %%CITATION = ARXIV:1207.1047;%%  
\bibitem{Emparan:2009cs} 
  R.~Emparan, T.~Harmark, V.~Niarchos and N.~A.~Obers,
  ``World-Volume Effective Theory for Higher-Dimensional Black Holes,''
  Phys.\ Rev.\ Lett.\  {\bf 102}, 191301 (2009)
  [arXiv:0902.0427 [hep-th]].
  %%CITATION = ARXIV:0902.0427;%%  
 %\cite{Emparan:2009at}
%\bibitem{Emparan:2009at} 
  R.~Emparan, T.~Harmark, V.~Niarchos and N.~A.~Obers,
  ``Essentials of Blackfold Dynamics,''
  JHEP {\bf 1003}, 063 (2010)
  [arXiv:0910.1601 [hep-th]].
  %%CITATION = ARXIV:0910.1601;%% 
%\cite{Grignani:2010xm}
%\bibitem{Grignani:2010xm} 
  G.~Grignani, T.~Harmark, A.~Marini, N.~A.~Obers and M.~Orselli,
  ``Heating up the BIon,''
  JHEP {\bf 1106}, 058 (2011)
  [arXiv:1012.1494 [hep-th]].
  %%CITATION = ARXIV:1012.1494;%%
%\cite{Grignani:2013ewa}
%\bibitem{Grignani:2013ewa} 
  G.~Grignani, T.~Harmark, A.~Marini and M.~Orselli,
  ``Thermal DBI action for the D3-brane at weak and strong coupling,''
  JHEP {\bf 1403}, 114 (2014)
  [arXiv:1311.3834 [hep-th]].
  %%CITATION = ARXIV:1311.3834;%%  
 \bibitem{Bigazzi:2009tc} 
  F.~Bigazzi, A.~L.~Cotrone and J.~Tarrio,
  ``Hydrodynamics of fundamental matter,''
  JHEP {\bf 1002}, 083 (2010)
  [arXiv:0912.3256 [hep-th]].
  %%CITATION = ARXIV:0912.3256;%%
 %\cite{Bigazzi:2010ku}
%\bibitem{Bigazzi:2010ku} 
  F.~Bigazzi and A.~L.~Cotrone,
  ``An elementary stringy estimate of transport coefficients of large temperature QCD,''
  JHEP {\bf 1008}, 128 (2010)
  [arXiv:1006.4634 [hep-ph]].
  %%CITATION = ARXIV:1006.4634;%% 
 %\cite{Tarrio:2013tta}
%\bibitem{Tarrio:2013tta} 
  J.~Tarrio,
  ``Transport properties of spacetime-filling branes,''
  JHEP {\bf 1404}, 042 (2014)
  [arXiv:1312.2902 [hep-th]].
  %%CITATION = ARXIV:1312.2902;%% 
 \bibitem{Kim:2013ysa} 
  N.~Kim,
  ``Holographic entanglement entropy of confining gauge theories with flavor,''
  Phys.\ Lett.\ B {\bf 720}, 232 (2013).
  %%CITATION = PHLTA,B720,232;%%
%\cite{Chang:2013mca}
%\bibitem{Chang:2013mca} 
  H.~C.~Chang and A.~Karch,
  ``Entanglement Entropy for Probe Branes,''
  JHEP {\bf 1401}, 180 (2014)
  [arXiv:1307.5325 [hep-th]].
  %%CITATION = ARXIV:1307.5325;%%  
%\bibitem{Kontoudi:2013rla} 
  K.~Kontoudi and G.~Policastro,
  ``Flavor corrections to the entanglement entropy,''
  JHEP {\bf 1401}, 043 (2014)
  [arXiv:1310.4549 [hep-th]].
  %%CITATION = ARXIV:1310.4549;%%  
%\bibitem{Karch:2014ufa} 
  A.~Karch and C.~F.~Uhlemann,
  ``Generalized gravitational entropy of probe branes: flavor entanglement holographically,''
  JHEP {\bf 1405}, 017 (2014)
  [arXiv:1402.4497 [hep-th]].
  %%CITATION = ARXIV:1402.4497;%%  
%\cite{Kol:2014nqa}
%\bibitem{Kol:2014nqa} 
  U.~Kol, C.~Nunez, D.~Schofield, J.~Sonnenschein and M.~Warschawski,
  ``Confinement, Phase Transitions and non-Locality in the Entanglement Entropy,''
  JHEP {\bf 1406}, 005 (2014)
  [arXiv:1403.2721 [hep-th]].
  %%CITATION = ARXIV:1403.2721;%%  
  %\cite{Chang:2014oia}
%\bibitem{Chang:2014oia} 
  H.~C.~Chang, A.~Karch and C.~F.~Uhlemann,
  ``Flavored N=4 SYM -- a highly entangled quantum liquid,''
  arXiv:1406.2705 [hep-th].
  %%CITATION = ARXIV:1406.2705;%%
\bibitem{Kumar:2012ui} 
  S.~P.~Kumar,
  ``Heavy quark density in N=4 SYM: from hedgehog to Lifshitz spacetimes,''
  JHEP {\bf 1208}, 155 (2012)
  [arXiv:1206.5140 [hep-th]].
  %%CITATION = ARXIV:1206.5140;%%
  %\cite{Faedo:2013aoa}
%\bibitem{Faedo:2013aoa} 
  A.~F.~Faedo, B.~Fraser and S.~P.~Kumar,
  ``Supersymmetric Lifshitz-like backgrounds from $\mathcal{N}$ = 4 SYM with heavy quark density,''
  JHEP {\bf 1402}, 066 (2014)
  [arXiv:1310.0206 [hep-th]].
  %%CITATION = ARXIV:1310.0206;%%    
%\cite{Bolognesi:2014dja}
%\bibitem{Bolognesi:2014dja} 
  S.~Bolognesi,
  ``Instanton Bags, High Density Holographic QCD and Chiral Symmetry Restoration,''
  arXiv:1406.0205 [hep-th].
  %%CITATION = ARXIV:1406.0205;%%
 
 
\end{thebibliography}
  \end{document}